\documentclass[a4paper,twocolumn,preprintnumbers,amsmath,amssymb,amsfonts,floatfix,nofootinbib,showkeys]{revtex4-1} %,linenumbers
\pdfoutput=1
\usepackage{dcolumn}
\usepackage{bm}
\usepackage{float}
\usepackage{epsfig}
\usepackage{graphicx}
\usepackage{longtable} 
\usepackage{rotating}
\usepackage{amsmath}
\usepackage{amsfonts}
\usepackage{amssymb}
\usepackage{xspace}
\usepackage{multirow}
\usepackage{mathrsfs}
\usepackage{calrsfs}
\usepackage{txfonts}
\usepackage{graphicx}
\usepackage{hypernat}
\usepackage{color}
\usepackage[utf8]{inputenc}
\usepackage[pdftex,colorlinks,linkcolor=red,citecolor=red,anchorcolor=red,urlcolor=red]{hyperref}
\ifpdf
\pdfoutput=1        % we are running pdflatex
\fi
\newif\ifarxiv
\arxivtrue
\newif\ifcomment
\newif\ifprint
\newif\ifnodef
\graphicspath{{./pics/}}
\newcommand {\version}     {\href{http://www.hepforge.org/downloads/tglaubermc}{v3.2}}
\newcommand{\pp}           {pp}
\newcommand{\np}           {np}
\newcommand{\pn}           {pn}
\newcommand{\nn}           {nn}
\newcommand{\AaAa}         {AA}
\newcommand{\AB}           {AB}
\newcommand{\NN}           {NN}
\newcommand{\pPb}          {pPb}
\newcommand{\pA}           {pA}

\newcommand{\PbPb}         {PbPb}
\newcommand{\CuCu}         {CuCu}
\newcommand{\AuAu}         {AuAu}
\newcommand{\XeXe}         {XeXe}

\newcommand{\ppbar}        {$\rm p\bar p$}
\newcommand{\pT}           {\ensuremath{p_{\rm T}}}
\newcommand{\RAA}          {\ensuremath{R_{\rm AA}}}
\newcommand{\TAA}          {\ensuremath{T_{\rm AA}}}
\newcommand{\TAB}          {\ensuremath{T_{_{\ensuremath{\it{AB}}}}}}
\newcommand{\AT}           {\ensuremath{A_{\rm T}}}
\newcommand{\TA}           {\ensuremath{T_{\rm A}}}
\newcommand{\TB}           {\ensuremath{T_{\rm B}}}
\newcommand{\Cent}         {\ensuremath{\mathcal{C}}}
\newcommand{\Lambdaqcd}    {\Lambda_{\ensuremath{\rm QCD}}}
\newcommand{\mean}[1]      {\ensuremath{\left<#1\right>}}
\newcommand{\rms}          {rms}
\newcommand{\eg}           {e.g.}
\newcommand{\ie}           {i.e.}
\newcommand{\cm}           {c.m.}
\newcommand{\dd}           {\ensuremath{{\rm d}}}
\newcommand{\dx}           {\ensuremath{{\rm d}x}}
\newcommand{\dy}           {\ensuremath{{\rm d}y}}
\newcommand{\dz}           {\ensuremath{{\rm d}z}}
\newcommand{\snn}          {\ensuremath{\sqrt{s_{\rm NN}}}}
\newcommand{\sqrts}        {\ensuremath{\sqrt{s}}}
\newcommand{\sigmaNN}      {\ensuremath{\sigma_{\rm NN}}}
\newcommand{\sigmaAB}      {\ensuremath{\sigma_{\rm AB}}}
\newcommand{\dmin}         {\ensuremath{d_{\rm min}}}
\newcommand{\dmax}         {\ensuremath{d_{\rm max}}}
\newcommand{\dnode}        {\ensuremath{d_{\rm node}}}
\newcommand{\bNN}          {\ensuremath{b_{\rm NN}}}

\newcommand{\Ncoll}        {\ensuremath{N_{\rm coll}}}
\newcommand{\Npart}        {\ensuremath{N_{\rm part}}}
\newcommand{\rcoll}        {\ensuremath{\rho_{\rm coll}}}
\newcommand{\rpart}        {\ensuremath{\rho_{\rm part}}}
\ifarxiv
\newcommand{\cldot}        {\ensuremath{\cdot}}
\else
\newcommand{\cldot}        {\ensuremath{\,}}
\fi

\newcommand{\hrefurl}[1]   {\href{#1}{\url{#1}}}
\newcommand{\Ref}[1]       {Ref.~\cite{#1}}

\newcommand{\Tab}[1]       {Table~\ref{#1}}
\newcommand{\Fig}[1]       {Fig.~\ref{#1}}
\newcommand{\Figs}[2]      {Figs.~\ref{#1} and \ref{#2}}
\newcommand{\Eq}[1]        {Eq.~\ref{#1}}
\newcommand{\Sec}[1]       {Sec.~\ref{#1}}

\newcommand{\Figure}[1]    {Figure~\ref{#1}}
\newcommand{\Figures}[2]   {Figures~\ref{#1} and \ref{#2}}
\newcommand{\Appendix}[1]  {Appendix~\ref{#1}}
\newcommand{\Section}[1]   {Section~\ref{#1}}

\newcommand{\MCG}          {MCG}
\newcommand{\ADND}         {ADND}
\newcommand{\co}[1]        {}

\begin{document}
%%%%%%%%%%%%%%%%%%%%%%%%%%%%%%%%%%%%%%%%%%%%%%%%%%%%%%%%%%%%%%%%%%%%%%%%%%%%%%%
\title{Improved Monte Carlo Glauber predictions at present and future nuclear colliders}
\date{\today}
\author{Constantin Loizides}\affiliation{ORNL, Oak Ridge, USA}
\author{Jason Kamin}\affiliation{UIC, Chicago, USA}
\author{David d'Enterria}\affiliation{CERN, Geneva, Switzerland\\}
%%%%%%%%%%%%%%%%%%%%%%%%%%%%%%%%%%%%%%%%%%%%%%%%%%%%%%%%%%%%%%%%%%%%%%%%%%%%%%%
\begin{abstract}\noindent
%%%%%%%%%%%%%%%%%%%%%%%%%%%%%%%%%%%%%%%%%%%%%%%%%%%%%%%%%%%%%%%%%%%%%%%%%%%%%%%
We present the results of an improved Monte Carlo Glauber (MCG) model of relevance for collisions involving nuclei at center-of-mass energies of BNL RHIC ($\snn=0.2$~TeV), CERN LHC~($\snn=2.76$--$8.8$~TeV), and proposed future hadron colliders~($\snn\approx 10$--$63$~TeV). 
The inelastic \pp\ cross sections as a function of $\snn$ are obtained from a precise data-driven parametrization that exploits the many available measurements at LHC collision energies.
We describe the nuclear density of a lead nucleus with two separated 2-parameter Fermi distributions for protons and neutrons to account for their different densities close to the nuclear periphery. 
Furthermore, we model the nucleon degrees of freedom inside the nucleus through a lattice with a minimum nodal separation, combined with a ``recentering and reweighting'' procedure, that overcomes some limitations of previous MCG approaches.
The  nuclear overlap function, number of participant nucleons and binary nucleon--nucleon collisions, participant eccentricity and triangularity, overlap area and average path length are presented in intervals of percentile centrality for lead--lead~(\PbPb) and proton--lead~(\pPb) collisions at all collision energies.
We demonstrate for collisions at $\snn=5.02$~TeV that the central values of the Glauber quantities change by up to 7\% in a few bins of reaction centrality, due to the improvements implemented, though typically remain within the previously assigned systematic uncertainties, while their new associated uncertainties are generally smaller (mostly below 5\%) at all centralities than for earlier calculations.
Tables for all quantities versus centrality at present and foreseen collision energies involving Pb nuclei, as well as for collisions of \XeXe\ at $\snn=5.44$~TeV, and \AuAu\ and \CuCu\ at $\snn=0.2$~TeV, are provided.
The source code for the improved Monte Carlo Glauber model is made publicly available.
\ifarxiv
\vspace{0.15cm}\\
\noindent {\bf Revisions \&\& changes of the arXiv document and code:}\\
 {\bf v1}, 19 Oct\, 2017: initial document, code v3.0 \\
 {\bf v2}, 24 May 2018: published version, code v3.1 includes fixes for spherical nuclei\\
 {\bf v3}, 15 Feb 2019: fixes tables in appendix with correct overlap calculation consistent with erratum,\\\color{white}{\bf v3}, 15 Feb 2019: \color{black}latest code \version\ includes changes from v2.7.
\fi
\end{abstract}
%\keywords{Glauber Monte Carlo model, number of binary nucleon--nucleon collisions, participant nucleons, nuclear overlap, participant eccentricity, participant triangularity, overlap area, average path length, centrality, TGlauberMC}
\maketitle
%%%%%%%%%%%%%%%%%%%%%%%%%%%%%%%%%%%%%%%%%%%%%%%%%%%%%%%%%%%%%%%%%%%%%%%%%%%%%%%
\section{Introduction}
\label{sec:intro}
%%%%%%%%%%%%%%%%%%%%%%%%%%%%%%%%%%%%%%%%%%%%%%%%%%%%%%%%%%%%%%%%%%%%%%%%%%%%%%%
The interpretation of many results measured in high-energy heavy ion collisions relies on the use of a model of the initial matter distribution resulting from the overlap of the two colliding nuclei at a given impact parameter $b$. 
Indeed, quantities such as (i)~the centrality dependence, expressed by the {\it number of participating nucleons} in the collision $\Npart(b)$, of any observable, 
(ii)~the nuclear {\it overlap function} $\TAA(b)$ or the {\it number of binary nucleon-nucleon collisions} $\Ncoll(b)$ used to derive the nuclear modification factor~($\RAA$) from the ratio of \AaAa\ over \pp\ spectra, (iii)~the elliptic and triangular flow parameters $v_2$ and $v_3$ normalized by the {\it eccentricity}  $\varepsilon_2(b)$ and {\it triangularity} $\varepsilon_3(b)$ of the overlap region, the average (iv)~surface area $\AT(b)$ and (v)~{\it path-length} $L(b)$ of the interaction region, all depend on a realistic model of the collision geometry~\cite{Miller:2007ri}. 

The standard method employed in high-energy heavy-ion collisions describes the initial transverse shape of the nuclei in terms of 2-parameter Fermi~(2pF) distributions~(also often called Wood-Saxon distributions) with half-density radius $R$ and diffusivity $a$ parameters obtained from fits to elastic lepton-nucleus data~\cite{DeJager:1974liz,DeJager:1987qc}, and determines the underlying multi-nucleon interactions in the overlap area between the nuclei through a Glauber eikonal approach~\cite{Glauber:1970jm}.
In the Monte Carlo Glauber~(\MCG) models~(\eg~\cite{Wang:1991hta,Alver:2008aq,Loizides:2014vua,Broniowski:2007nz,Rybczynski:2013yba,Loizides:2016djv}), individual nucleons are sampled event-by-event from the underlying 2pF distributions and the collision properties are calculated by averaging over multiple events.
However, neutron-rich nuclei, such as $^{208}$Pb may have differing proton and neutron density distributions at the nuclear periphery.
Indeed, measurements have recently been able to extract the neutron profile of several nuclei that show differences with respect to their proton distribution~\cite{Klos:2007is,Tarbert:2013jze}, and various works have already studied its impact on different isospin--dependent observables in nuclear collisions~\cite{Paukkunen:2015bwa,De:2016ggl,Helenius:2016dsk}.
%It is known, however, that the proton and neutron distributions may not be exactly the  same at the surface of heavy stable nuclei~\cite{barrett1977nuclear}. 
%Such an effect is particularly important in neutron-rich nuclei, such as $^{208}$Pb with an $N/Z\approx1.5$ neutron excess, where the large Coulomb barrier reduces the proton density at the surface of the nucleus and concurrently ``pushes out'' the neutrons against the surface tension generated by the nuclear mean-field.
%Such an  effect is particularly large in neutron-rich nuclei, such as $^{208}$Pb with an $N/Z\approx1.5$ neutron excess, where the excess of neutrons in the nuclear periphery enhances the surface tension reducing the countereffect of the proton Coulomb repulsion, and hence effectively increases the binding energy of the nucleus. %the large Coulomb barrier reduces the proton density at the surface of the nucleus and ``pushes'' neutrons out to the nuclear periphery.
%As a result, the neutron can have either a ``skin-type'' distribution with 2pF half-density radius larger than that of the proton~($R_n>R_p$) but equal diffusivity parameter~($a_n=a_p$)~\cite{Tsang:2012se,Tarbert:2013jze,Zenihiro:2010zz}, 
%or a ``halo-type'' distribution with $R_n=R_p$ and $a_n>a_p$~\cite{Trzcinska:2001sy}.

In this article, we present the results of improved Glauber Monte Carlo model calculations for $\Ncoll(b)$, $\Npart(b)$, $\TAA(b)$, $\varepsilon_2(b)$, $\varepsilon_3(b)$, $\AT(b)$ and $L(b)$ in \PbPb\ and \pPb\ collisions at LHC~($\snn=2.76$, $5.02$, $5.5$, $8.16$ and $8.8$~TeV), High-Energy LHC ($\snn=10.6$, $17$~TeV), and Future Circular Collider FCC~($\snn=39$ and $63$~TeV)~\cite{Dainese:2016gch} energies, by considering for the first time separated transverse profiles for protons and neutrons in the lead nucleus. 
The corresponding values for the inelastic \pp\ cross section are obtained from a data-driven parametrization with reduced uncertainties thanks to the many available measurements at LHC collision energies.
The nucleon degrees of freedom inside a nucleus are modeled using a lattice with a minimum nodal separation, that mimics hard-core repulsion between nucleons without distorting the nuclear density.
Residual small distortions in the generated nuclear densities resulting from adjusting the nucleon center-of-mass (\cm) with that of the nucleus are overcome by reweighting the original nuclear density.
We exemplify for collisions at $\snn=5.02$~TeV that the central values of $\Ncoll(b)$, $\Npart(b)$, $\TAA(b)$, and $\varepsilon_2(b)$ change due to the inclusion of the separated proton and neutron transverse distributions, but typically remain within the previously assigned systematic uncertainties.
Their new associated uncertainties are generally smaller than for earlier calculations except for mid-peripheral events where they are slightly larger in some cases.
Tables for all quantities versus centrality at present and foreseen collision energies involving Pb nuclei are provided. 
Results for other studied systems, such as \AuAu\ and \CuCu\ collisions at $\snn=0.2$~TeV and \XeXe\ collisions at $\snn=5.44$~TeV, are provided also for completeness.
As for previous versions of the model, the source code for ``TGlauberMC'' (\version) has been made publicly available at HepForge~\cite{glaucode}.

The paper is organized as follows:
\Section{sec:glauber} describes the basic quantities of interest computed in the article. 
\Section{sec:sigmaNN} presents a parametrization of the \cm\ energy dependence of the nucleon inelastic cross section~($\sigmaNN$) based on existing proton--proton (\pp) and proton--antiproton (\ppbar) data.
\Section{sec:details} introduces the basic details of the \MCG\ calculation.
\Section{sec:improvements} discusses the improvements of the \MCG\ modeling, namely using a more realistic nuclear matter density with separated protons and neutrons profiles~(\Sec{sec:nstd}), incorporating a minimum inter-nucleon separation without distorting the nuclear profile~(\Sec{sec:mns}), reweighting the nuclear density to compensate residual distortions introduced by the nucleon center-of-mass recentering~(\Sec{sec:rec}), and using a more precise parameterization of the $\sigmaNN$~(\Sec{sec:nnmod}).
\Section{sec:results} presents the results of the improved \MCG\ calculation and \Section{sec:summary} summarizes our main conclusions.
\Appendix{app:opt} illustrates the difference between an optical and a Monte Carlo Glauber calculation.
\Appendix{app:nnmod} briefly discusses the inclusion of subnucleonic degrees of freedom in the \MCG\ calculation.
\Appendix{app:guide} provides an updated user's guide for running the publicly available \MCG\ code.
\Appendix{app:tab} provides tables with calculated quantities for all relevant collision energies involving Pb nuclei, including \XeXe\ collisions at $\snn=5.44$~TeV as well as \AuAu\ and \CuCu\ collisions at $\snn=0.2$~TeV.

%%%%%%%%%%%%%%%%%%%%%%%%%%%%%%%%%%%%%%%%%%%%%%%%%%%%% 
\section{Glauber formalism}
\label{sec:glauber}
%%%%%%%%%%%%%%%%%%%%%%%%%%%%%%%%%%%%%%%%%%%%%%%%%%%%% 
The standard procedure to determine the transverse overlap area, and other derived quantities in a generic proton--nucleus~(\pA) or nucleus--nucleus collision~(\AB) at impact parameter $b$, is based on a simple Glauber multi-scattering eikonal model that assumes straight-line trajectories of the nucleons from the two colliding nuclei~\cite{Glauber:1970jm}.
A review that describes the basic formalism can be found in~\cite{Miller:2007ri}, of which we briefly summarize the main concepts here.

%%%First, we briefly summarize the main concepts of the Glauber formalism. [redundant with sentence above]
To simplify the mathematical description, the reaction plane of the two colliding nuclei, i.e.\ the plane defined by the impact parameter and the beam direction, is given by the $x$- and $z$-axes, while the transverse plane is given by the $x$- and $y$-axes.
The collision impact parameter~$b$ is distributed assuming ${\rm d}N/{\rm d}b \propto b$, and the centers of the nuclei are shifted to $(-\frac{b}{2},0,0)$  and $(\frac{b}{2},0,0)$.

In ``optical'' Glauber calculations a smooth nuclear matter density, $\rho$, for each nucleus is used and properties of the reaction zone and all derived quantities are analytically calculated.
In Monte Carlo based approaches individual nucleons are distributed for each nucleus according to $\rho$ in an event-by-event basis and collision properties as well as derived quantities are calculated by averaging over multiple events. 
In both cases, following the eikonal ansatz, the nucleons are assumed to move in straight trajectories along the beam axis.
The nuclear reaction is modeled by successive independent interactions between two nucleons from different nuclei, where the interaction strength between two nucleons is typically modeled using the nucleon--nucleon inelastic cross section~($\sigmaNN$) in the transverse plane.
The transverse positions of nucleons are assumed to be constant during the short passage time of the two high-energy nuclei, while their longitudinal coordinate does not play a role in the calculation.

The optical calculations are based on the {\it thickness function} of a nucleus which quantifies the transverse nucleon density as $T(x,y)=\int\,\rho(x,y,z)\,\dz$, which is usually normalized to the number of nucleons in the nucleus A.
The {\it nuclear overlap function} of nuclei $A$ and $B$ colliding at impact parameter $b$, $\TAB(b)$, can then be expressed as the convolution of the corresponding thickness functions of $A$ and $B$
\begin{eqnarray}
\TAB(b) &=& \int \rcoll(x,y,b) \, \dx\dy \nonumber\\
        &=& \int \TA\left(x-\frac{b}{2},y\right)\TB\left(x+\frac{b}{2},y\right)\,\dx\dy
\label{eq:nuc_overlap}
\end{eqnarray}
usually normalized so that $\int \TAB(b)\,b\,\dd b=AB$.
%In {\it minimum bias} reactions, the {\it average} nuclear thickness and nuclear overlap functions are $\langle \TA\rangle_{\rm MB} \equiv \frac{\int \dd^2b \, \TA}{\int \dd^2b} = \frac{A}{\pi\,R_{\rm A}^2}=\frac{A}{\sigma_{\rm pA}^{\rm geo}}$ and $\langle \TAB\rangle_{\rm AB} \equiv \frac{\int \dd^2b \, \TAB(b)}{\int \dd^2b} = \frac{AB}{\pi (R_{\rm A}+R_{\rm B})^2}=\frac{AB}{\sigmaAB^{\rm geo}}$.

The number of nucleons in the target and projectile nuclei that interacted at least once in a collision at impact parameter $b$ is called the number of participants~(or ``wounded nucleons''), and calculated as~\cite{Bialas:1976ed,Kharzeev:1996yx}
\begin{eqnarray}
\Npart(b) &=& \int {\rpart(x,y,b) \, \dx\dy} \nonumber\\
          &=&A\int {\TA^-\left({1 - \left[ {1 - \sigmaNN\TB^+}\right]^B }\right)\dx\dy} \nonumber\\
          &+&B\int {\TB^+\left({1 - \left[ {1 - \sigmaNN\TA^-}\right]^A }\right)\dx\dy}          
\label{eq:Npart}
\end{eqnarray}
with $T_{\rm X}^{\pm}\equiv T_{\rm X}(x\pm\frac{b}{2},y)$.
Similarly, the total number of binary nucleon--nucleon collisions at impact parameter $b$ is calculated as 
\begin{eqnarray}
\Ncoll(b) &=& \sigmaNN\,\int \rcoll(x,y,b)\,\dx\dy \nonumber\\
          &=& \sigmaNN \TAB(b)\,.
\label{eq:Ncoll}
\end{eqnarray}
Hence, the nuclear overlap function, $\TAB(b)=\Ncoll(b)/\sigmaNN$, can be thought of as the nucleon-nucleon luminosity~(reaction rate per unit cross section) in an \AB\ collision at a given impact parameter $b$.

\MCG\ calculations obtain the quantities~(\ref{eq:Npart}) and (\ref{eq:Ncoll}) by simply counting either the number of nucleons that interacted at least once~(\Npart), or the total number of individual nucleon--nucleon collisions~($\Ncoll$), where the collisions between the nucleons of the two incoming nuclei are determined by a $\sigmaNN$-dependent interaction probability in the transverse plane.

The second moment, also called eccentricity~\cite{Alver:2006wh}, the third moment, also called triangularity~\cite{Alver:2010gr}, and higher moments~\cite{Teaney:2010vd} of the collision region at impact parameter $b$, which are used to characterize the initial geometrical shape, are given by
\begin{equation}
\varepsilon_{n}(b)=\frac{\left<r^n\cos(n\phi-n\psi)\right>}{\left<r^n\right>}
\label{eq:ecc}
\end{equation}
where $n$ denotes the moment~($n=2$ for eccentricity, $n=3$ for triangularity), $r=\sqrt{x^2+y^2}$ and $\psi=\tan^{-1}\frac{y}{x}$.
The averages are performed by considering the central positions of either, participant nucleons or binary nucleon--nucleon collisions, or of an admixture of the two.

The effective transverse overlap area between the two nuclei is often taken to be proportional to the widths of the participant distributions
\begin{equation}
A_{\perp}(b) \propto \sqrt{\mean{x^2}\mean{y^2}}
\label{eq:A_perp} 
\end{equation}
where the averages are taken over participant nucleons.  
There is no commonly accepted definition of the absolute normalization of the overlap area.
Historically, either $\pi$~\cite{Alver:2008zza} or $4\pi$~\cite{Drescher:2007cd} have been used, where the latter essentially coincides with the geometrical overlap area of two uniform disks. Recently, it was proposed to directly calculate the area in the \MCG\ by evaluating the area of wounded nucleons with a fine-grained grid~\cite{Loizides:2016djv}.

The average path-length through a static medium with a density parameterized with $\rpart$ can be calculated using
\begin{equation}
L(b)= \frac{\int l\,\rpart(x_0+l\cos\phi_0,y_0+l\sin\phi_0,b)\,\dd l\,\dd P_0}{0.5\int \rpart(x_0+l\cos\phi_0,y_0+l\sin\phi_0,b)\,\dd l\,\dd P_0}
\label{eq:L} 
\end{equation}
where the initial point $(x_0,y_0)$ is usually distributed according to $\rcoll$, and the azimuthal direction $\phi_0$ uniformly~\cite{Drees:2003zh,Dainese:2004te}.

The total inelastic cross sections for \pA\ or \AB\ collisions are
\begin{equation}
\sigma_{\rm pA}=2\pi\,\int b\, \left[1-e^{-\sigmaNN\,\TA(b)}\right]\, \dd b\;,
\label{eq:glauber_spA}
\end{equation}
and
\begin{equation}
\sigmaAB=2\pi\,\int b\, \left[1-e^{-\sigmaNN\,\TAB(b)}\right]\, \dd b\,.
\label{eq:glauber_sAA}
\end{equation}
\MCG\ calculations obtain the cross sections by simply multiplying the fraction of accepted events with $\pi b^2_{\rm max}$, where $b_{\rm max}$ is the maximum generated impact parameter~(usually $20$~fm). 

Observables are often studied in intervals of cross sections, called ``centrality percentiles'', whose experimental ranges are typically obtained by ordering the events according to their particle multiplicity or transverse energy and, in the case of \AaAa\ collisions, can be translated into equivalent ranges of impact parameter~(see \eg\ \cite{Abelev:2013qoq}).
Instead of reporting results as a function of centrality percentiles, often the mean number of participants in the centrality interval is used, which, like all quantities in a Glauber calculation, can be obtained by performing the calculation over their respective impact parameter range.
\ifcomment 
Experimentally one uses the ``reaction centrality'' $\Cent$ as a proxy for the impact-parameter $b$ of a given nucleus-nucleus collision, by dividing the  particle production cross section into centrality bins $\Cent_k = \Cent_1, \Cent_2, \cdots $ according to some fractional interval $\Delta \Cent$ of the total cross section, \eg\ $\Delta \Cent = 0.0 - 0.1$ represents the 10\% most central collisions. 
The fraction of the geometric cross section within impact parameters $b_{1}<b<b_{2}$ is
\begin{equation}
f_{\rm geo}(b_1<b<b_2)\;=\; 2\pi\; \int_{b_1}^{b_2} b\, \dd b \left(1-e^{-\sigmaNN \TAB(b)}\right) / \sigmaAB \; ,
\label{eq:f_geo}
\end{equation}
which simply corresponds to a given centrality percentile $\Delta \Cent$ between 0\% (most central) and 100\% (most peripheral).
\fi

\begin{figure}[t!]
\centering
\includegraphics[width=0.5\textwidth]{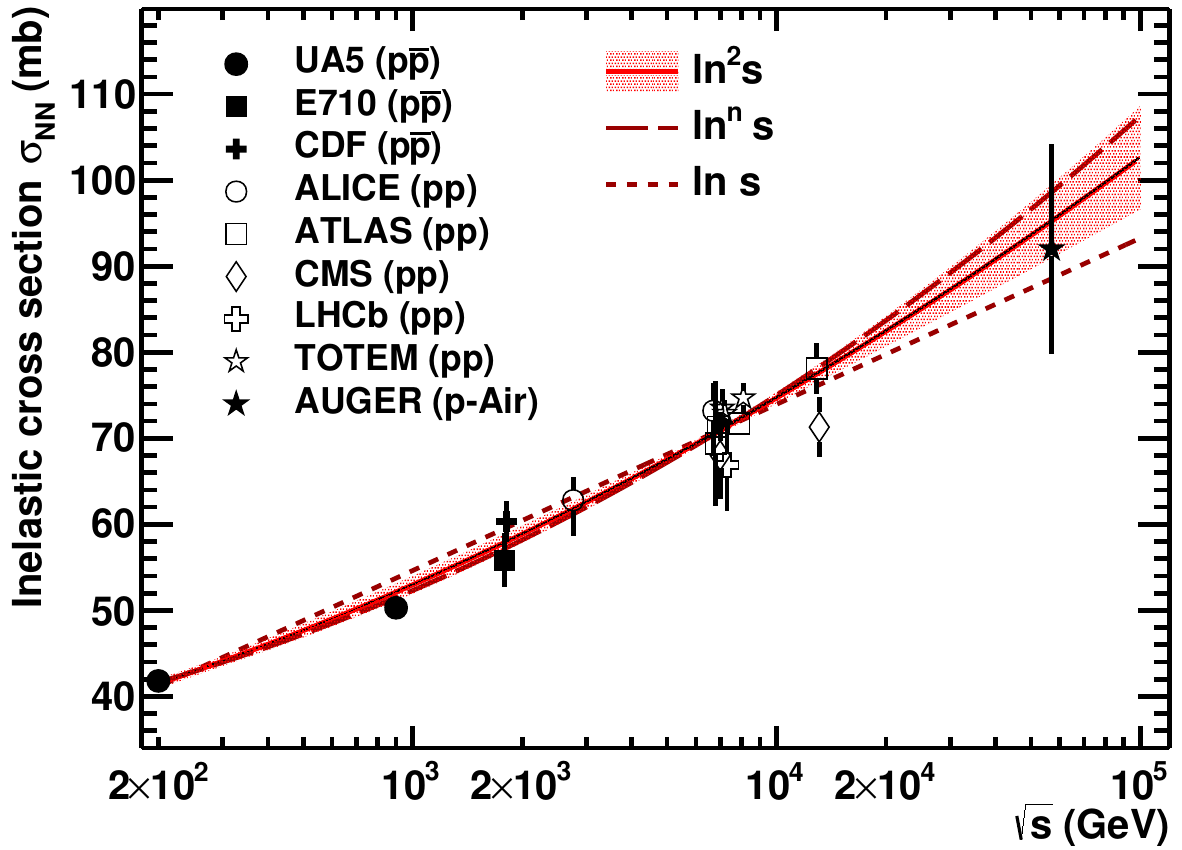} %from cs/sigma_inel_pp_vs_sqrts_plot.C
\caption{Inelastic \pp\ cross section as a function of \cm\ energy in the range $\sqrts=0.2$--$100$~TeV. 
         Experimental data points at various colliders and cosmic-ray energies from UA5~\cite{Alner:1986iy}, E710~\cite{Amos:1990jh,Amos:1991bp}, CDF~\cite{Abe:1993xy,Abe:1993xx}, ALICE~\cite{Abelev:2012sea}, ATLAS~\cite{Aad:2011eu,Aad:2014dca,Aaboud:2016ijx,Aaboud:2016mmw}, CMS~\cite{Chatrchyan:2012nj,VanHaevermaet:2016gnh}, LHCb~\cite{Aaij:2014vfa}, TOTEM ~\cite{Antchev:2011vs,Antchev:2013iaa,Antchev:2013paa} and AUGER~\cite{Collaboration:2012wt}.
         Fits of $\ln s$, $\ln^2 s$ and $\ln^{n} s$ to the data are shown~(for details see text).}
\label{fig:sigmappvs}
\end{figure}
\begin{table}[t!]
\begin{tabular}{l|c|c|c|c}\hline\hline
\centering
 Type              & $A$            & $B$                & $n$               & $\chi^2/N_{\rm dof}$ \\\hline
 $\ln s$           & $-3.33\pm1.58$ & $4.195\pm0.103$    & $1$~(fixed)       & $1.52$ \\
 $\ln^2 s$         & $25.0\pm0.9$   & $0.146\pm0.004$    & $2$~(fixed)       & $0.97$ \\
 $\ln^n s$         & $29.8\pm4.7$   & $0.038\pm0.060$    & $2.43\pm0.50$     & $0.98$ \\
\hline\hline
\end{tabular} 
\caption{\label{tab:snnfitvals}Fit values and $\chi^2/N_{\rm dof}$ for the collision-energy dependence of $\sigmaNN$ parameterized by \Eq{eq:signnfunc} and displayed in \Fig{fig:sigmappvs}.}
\end{table}

%%%%%%%%%%%%%%%%%%%%%%%%%%%%%%%%%%%%%%%%%%%%%%%%%%%%% 
\section{Parameterization of the inelastic nucleon--nucleon cross section}
\label{sec:sigmaNN}
%%%%%%%%%%%%%%%%%%%%%%%%%%%%%%%%%%%%%%%%%%%%%%%%%%%%% 
A fundamental ingredient of any Glauber calculation is the inelastic nucleon-nucleon cross section, $\sigmaNN$, at the same \cm\ energy $\snn$ of the nuclear collision under consideration. 
The value of $\sigmaNN$ includes particle production contributions from both (semi)hard parton-parton scatterings, computable above a given $\pT\approx$~2~GeV cutoff by perturbative QCD approaches, as well as from softer ``peripheral'' scatterings of diffractive nature, with a scale not very far from $\Lambdaqcd\approx0.2$~GeV. 
%In addition, diffractive particle production (from single and double diffractive scatterings as well as double-Pomeron exchange) constitutes also a significant fraction, up to 30\%~\cite{Ostapchenko:2010gt}, of $\sigmaNN$ at the collider energies of interest here. 
Today, $\sigmaNN$ cannot be computed from first-principle QCD calculations~(although future developments in lattice QCD computations could improve this situation) and one resorts to phenomenological approaches to fit the experimental data and predict their evolution as a function of $\snn$~\cite{dEnterria:2016oxo}.
At high \cm\ energies, above a few tens of GeV, \pp\ and \ppbar~(as well as \nn\ and \np) collisions all feature the same inelastic cross sections.  
Any potential differences due to their different valence-quark structure are increasingly irrelevant, and all existing experimental measurements can be combined to extract $\sigmaNN$. 
The $\sqrts$ dependence of the inelastic cross section $\sigmaNN$ is shown in \Fig{fig:sigmappvs} for all the available data from \ppbar\ and \pp\ colliders, and the AUGER result at $\sqrts=57$~TeV derived from cosmic-ray data~\cite{Collaboration:2012wt}. 
We include \ppbar\ measurements from UA5~\cite{Alner:1986iy} at $\sqrts=200$ and $900$~GeV, E710~\cite{Amos:1990jh,Amos:1991bp} and CDF~\cite{Abe:1993xy,Abe:1993xx} at $\sqrts=1.8$~TeV, as well as \pp\ results from ALICE at 7 TeV~\cite{Abelev:2012sea}, ATLAS at $7$, $8$ and $13$~TeV~\cite{Aad:2011eu,Aad:2014dca,Aaboud:2016ijx,Aaboud:2016mmw}, CMS at $7$ and $13$~TeV~\cite{Chatrchyan:2012nj,VanHaevermaet:2016gnh}, LHCb at $7$ TeV~\cite{Aaij:2014vfa}, and TOTEM at $7$ and $8$ TeV~\cite{Antchev:2011vs,Antchev:2013iaa,Antchev:2013paa}. 
The experimental $\sigmaNN$ values plotted are either obtained (i) from the subtraction $\sigma_{\rm inel} = \sigma_{\rm tot} - \sigma_{\rm el}$, where $\sigma_{\rm tot}$ and $\sigma_{\rm el}$ have been accurately measured in dedicated forward Roman pot detectors (TOTEM~\cite{Antchev:2011vs,Antchev:2013iaa,Antchev:2013paa} and ALFA~\cite{Aad:2014dca,Aaboud:2016ijx,Aaboud:2016mmw}), or (ii) from measurements of inelastic particle production data in the central detectors collected with ``minimum bias'' triggers.
The latter measurements are less accurate than the former, as they require an extrapolation, dominated by diffractive contributions, to forward regions of phase space not covered by detectors, and therefore have larger uncertainties.

\begin{table}[t]
\begin{tabular}{c|c}\hline\hline
\centering
 $\sqrts$ (TeV) & $\sigmaNN$ (mb) \\\hline
 0.2            & $41.6\pm0.6$ \\
 0.9            & $52.2\pm1.0$ \\
 2.76           & $61.8\pm0.9$ \\
 5.02           & $67.6\pm0.6$ \\
 5.44           & $68.4\pm0.5$ \\
 5.5            & $68.5\pm0.5$ \\
 7              & $70.9\pm0.4$ \\
 8              & $72.3\pm0.5$ \\
 8.16           & $72.5\pm0.5$ \\
 8.8            & $73.3\pm0.6$ \\
 10.6           & $75.3\pm0.7$ \\
 13             & $77.6\pm1.0$ \\
 14             & $78.4\pm1.1$ \\
 17             & $80.6\pm1.5$ \\
 27             & $86.0\pm2.4$ \\
 39             & $90.5\pm3.3$ \\
 63             & $96.5\pm4.6$ \\
 100            & $102.6\pm6.0$\\
\hline\hline
\end{tabular} 
\caption{\label{tab:signnvalues} Values of the nucleon-nucleon inelastic cross section $\sigmaNN$ extracted from the $\ln^2$ fit, with the uncertainties estimated from the difference of the $\ln s$ and $\ln^n s$ ($n=2.43$) fits at collision-energies $\sqrts$ relevant for RHIC, LHC, and FCC.}
\end{table}

The collision-energy dependence of $\sigmaNN$ has been fit to the parameterization
\begin{equation}
\sigmaNN(s) = A + B \,\ln^{n}(s)\,,
\label{eq:signnfunc}
\end{equation}
where $n$ was fixed to either $n=1$ or $n=2$, or otherwise left free in the fit.
The values and $\chi^2/N_{\rm dof}$ for the three cases are given in \Tab{tab:snnfitvals}.
The $n=2$ case, which represents the asymptotic $\sqrts$-dependence expected to saturate the Froissart bound~\cite{Froissart:1961ux}, is used as central value for the interpolation (and extrapolation) of $\sigmaNN$ versus $\sqrts$, listed in \Tab{tab:signnvalues} for relevant LHC and FCC energies.
The difference~(normalized by 2.4 to account for the full width at half maximum) of the so-derived $\sigmaNN$ values to those obtained for $n=1,2.43$ is assigned as systematic uncertainty (shown as a band in \Fig{fig:sigmappvs}).
The resulting cross section at 100 TeV of $\sigmaNN=102.6\pm6.0$~mb is consistent with the value $105.1\pm2.0$~mb, obtained from the average of various model calculations~\cite{dEnterria:2016oxo}.
The value extracted for the top RHIC energy of $\snn=0.2$~TeV is $41.6 \pm 0.6$~mb and is consistent with the typically used value of $42\pm3$~mb~\cite{Alver:2008zza}.

Other more complicated functional forms were also tried, motivated by the Ansatz used by the COMPETE collaboration~\cite{Cudell:2002xe}, such as $\sigmaNN(\sqrts) = A + B \,\ln^{2}(s) + C\, s^{-\eta}$, and $\sigmaNN(\sqrts) = A + B \,\ln^{2}(s) + D \ln(s)$.
The corresponding fits resulted in $A=24.4\pm1.4$, $B=0.1008\pm0.1537$, $C=1.454\pm1.768$, $\eta=0.131\pm0.0180$ with $\chi^2/N_{\rm dof}=1.09$, and $A=39.7\pm1.4$, $B=0.2212\pm0.0708$, $D=-2.154\pm2.035$ with $\chi^2/N_{\rm dof}=0.96$, respectively.
Both parameterizations turned out to be numerically close to \Eq{eq:signnfunc} for $n\approx2.43$ as determined by the simpler $\ln^n s$-fit.  

From the obtained values of $\sigmaNN$, one can then easily derive the corresponding proton-nucleus and nucleus-nucleus inelastic collisions making use of \Eq{eq:glauber_spA} and \Eq{eq:glauber_sAA}. 
The computed $\sigma_{\rm pA}$ and $\sigmaAB$ results for all relevant collision systems in this work are listed in \Tab{tab:ssum}.
The Glauber calculation gives $\sigma_{\rm\PbPb}^{\rm MC}=7.57\pm0.03$~b and $\sigma_{\rm\pPb}^{\rm MC}=2.08\pm0.01$~b for the hadronic \PbPb\ and \pPb\ cross sections, in good agreement with the measured values of $\sigma_{\rm\PbPb}=7.7\pm0.6$~b at $\snn=2.76$~TeV~\cite{ALICE:2012aa} and $\sigma_{\pPb}=2.06\pm0.08$~b~\cite{Khachatryan:2015zaa} as well as $\sigma = 2.10 \pm 0.07$~\cite{Abelev:2014epa} at $\snn=5.02$~TeV, respectively.

%%%%%%%%%%%%%%%%%%%%%%%%%%%%%%%%%%%%%%%%%%%%%%%%%%%%%%%%%%%%%%%%%%%%%%%%%%%%%%%
\section{Details of the MCG calculation}
\label{sec:details}
%%%%%%%%%%%%%%%%%%%%%%%%%%%%%%%%%%%%%%%%%%%%%%%%%%%%%%%%%%%%%%%%%%%%%%%%%%%%%%%
The implementation of the \MCG\ calculation is described in detail in \cite{Alver:2008aq,Loizides:2014vua}.
It consists of two steps: first, \textit{constructing} the nuclei and, second, \textit{colliding} the nuclei. 

To construct a nucleus, the position of each nucleon is determined according to a probability density function usually taken from measurements of the charge density distribution of the nucleus~\cite{DeJager:1974liz,DeJager:1987qc}.
%For simplicity, one takes the same spatial distribution for all parton species in the proton (valence and sea quarks, gluons) and momenta, and neglects any correlations among them.
For spherical nuclei, the nucleon positions can be determined in polar coordinates with a uniform distribution for the azimuthal and polar angles, coupled with a 2pF distribution in the radial direction
\begin{equation} 
\rho(r) = \frac{\rho_{0}}{1+ \exp{[(r-R)/a]}} \,,
\label{eq:2pF}
\end{equation} 
where $\rho_0$ is a normalization constant so that $\int d^3 r \,\rho(r)=1$.
The half-density or central radius $R$ describes the mean location of the nucleus area~(\ie\ $R$ is indicative of the extension of the bulk part of the density distribution).
The diffusivity parameter $a$ describes the tail of the density profile. 
%The associated $rms$ radius is $R_{rms}= \sqrt{3/5}R$ in the limit $a/R \to 0$ and  $R_{rms}=\sqrt{12}a$ for $a/R \to \infty$. For $a/R$~=~0.2, $R_{rms}~\approx~1.07\,R$.
%The nuclear radius follows roughly a mass-number $A$ dependence of the type $R_A = 1.19\cdot A^{1/3} - 1.61\cdot A^{-1/3}$ fm and the surface thickness $a=0.54$ fm.
%The nuclear density parameters used for the 2pF distributions are typically taken from the Atomic Data and Nuclear Data Tables~(\ADND)~\cite{DeJager:1987qc,DeJager:1974liz,Fricke:1995zz,Angeli:2004kvy,Angeli:2013epw}.
%These values are extracted via coulomb scattering in electron-nucleus and muon-nucleus scattering experiments and therefore probe the \textit{charge} density of the nucleus.  
Values for Pb nuclei are listed in \Tab{tab:awR}, while a complete list of parameters for other nuclei can be found in \Ref{Loizides:2014vua}.
To mimic a hard-core repulsion potential between nucleons, a minimum inter-nucleon separation~($\dmin$) of usually $0.4$~fm between their centers is enforced when sampling the positions of the nucleons inside a nucleus.
In order to ensure that the center-of-mass of each constructed nucleus is at $(0,0,0)$, the nucleons are individually ``recentered'' through a procedure discussed in more detail later. 

To simulate the collision, the centers of the nuclei are then shifted to $(-b/2,0,0)$  and $(b/2,0,0)$. 
The collision of two nuclei is then modeled by assuming that the nucleons of each nucleus travel in a straight line along the beam axis in the transverse plane (eikonal approximation), ignoring their longitudinal coordinates in the calculation.
The impact parameter of the collision is chosen randomly from ${\rm d}N/{\rm d}b \propto b$ up to some large maximum $b_{\rm max}\simeq 20\,$fm, chosen to be significantly greater than twice the nuclear radius. 
Two nucleons from different nuclei are usually assumed to collide if their relative transverse distance is less than a diameter given by
\begin{equation}
  D = \sqrt{\sigmaNN/\pi}
  \label{eq:mc_collisions}
\end{equation}
which geometrically parameterizes the interaction strength of two nucleons for a given value of $\sigmaNN$.
If no nucleon--nucleon collision is registered for any pair of nucleons, then no nucleus--nucleus collision occurred. 
Counters for determination of the total~(geometric) cross section are updated accordingly.
The inelastic nucleon-nucleon cross section $\sigmaNN$ is either directly taken from measurements in \pp\ collisions, or extracted from interpolations of the available data, as explained in \Sec{sec:sigmaNN}.

Constructing the nucleus is a principal ingredient of the \MCG\ model and the dominant source of systematic uncertainties in the Glauber quantities, in particular after reducing the uncertainties of the interpolated \sigmaNN\ values. %discussed in \Sec{sec:sigmaNN}.
In the following, we will discuss improvements of the \MCG\ model aiming at achieving a more accurate baseline description with reduced systematic uncertainties. 
The new results will be labeled {\it improved} \MCG, and discussed in detail in the next Section. 
In order to compare with previous baseline results, we compare the results of our new calculations with those from a set of {\it traditional} parameters of the \MCG\ model, typically used in previous studies~\cite{Abelev:2013qoq,Adam:2014qja,Loizides:2016djv}, given by $\sigmaNN=64$~mb for $\snn=2.76$~TeV and $\sigmaNN=70$~mb for $\snn=5.02$~TeV with an uncertainty of $\pm5$~mb, with charge radius and diffusivity of the nuclear density profile varied within their measured $1\sigma$ uncertainties, and minimum inter-nucleon separation distance varied by 100\%, \ie\ between $0$ and $0.8$~fm.
The algorithmic definitions, as well as central values and uncertainties of the parameters, for the {\it traditional} and {\it improved} \MCG\ setups are summarized in \Tab{tab:settings}.

\begin{table}[t!] 
  \begin{tabular}{l|c|c}\hline\hline 
\centering
  \MCG\ model                & Traditional         & Improved \\\hline
  Density for Pb             & Charge, 2pF (``Pb'')     & Point, D2pF (``Pbpnrw'')                 \\
  NN separation (fm)~~~      & $\dmin=0.4\pm0.4$        & $\dnode=0.4\pm0.4$ ~~                    \\
  $\sigmaNN$ (mb)~~~         & $70\pm5$                 & $67.6\pm0.5$                             \\
  Recentering                & Shift                    & $\dmax=0.1$~fm + reweight                \\\hline
  TGlauberMC                 & $\equiv$v2.x             & $\equiv$v3.x                             \\
\hline\hline 
\end{tabular} 
\caption{Parameters with corresponding uncertainties for the traditional and improved \MCG\ model used to compute Glauber quantities for nuclear collisions at a reference \cm\ energy of $\snn=5.02$~TeV.}
\label{tab:settings} 
\end{table} 

%In the case of \pPb\ collisions, the detailed shape of the proton distribution is not relevant.
%We use an exponential distribution $\exp\left(-r/R\right)$ with $R=0.234$~fm based on the measured form factor of the proton~\cite{Hofstadter:1956qs}. %, and uniformly distributed for the solid angle.

%%%%%%%%%%%%%%%%%%%%%%%%%%%%%%%%%%%%%%%%%%%%%%%%%%%%%%%%%%%%%%%%%%%%%%%%%%%%%%%
\section{Improvements of the \MCG\ modeling}
\label{sec:improvements}
%%%%%%%%%%%%%%%%%%%%%%%%%%%%%%%%%%%%%%%%%%%%%%%%%%%%%%%%%%%%%%%%%%%%%%%%%%%%%%%
%%%%%%%%%%%%%%%%%%%%%%%%%%%%%%%%%%%%%%%%%%%%%%%%%%%%% 
\subsection{Nuclear matter density}
\label{sec:nstd}
%%%%%%%%%%%%%%%%%%%%%%%%%%%%%%%%%%%%%%%%%%%%%%%%%%%%% 
The nuclear density parameters used for the 2pF distributions are typically taken from the Atomic Data and Nuclear Data Tables~(\ADND)~\cite{DeJager:1987qc,DeJager:1974liz,Fricke:1995zz,Angeli:2004kvy,Angeli:2013epw}.
They are extracted via Coulomb scattering in electron--nucleus and muon--nucleus measurements and therefore dominantly probe the \textit{charge} density of the nucleus.
%Electromagnetic nuclear interactions~\cite{DeJager:1974liz,DeJager:1987qc}, from which the nucleon parameters of the top rows of \Tab{tab:awR} are extracted, are mostly sensitive to the {\it charged} (\ie\ proton) distribution in the nucleus, and only indirectly to the neutron distribution via the neutron magnetic moment. 
%The nuclear density parameters used for the 2pF distributions are extracted via coulomb scattering in electron--nucleus and muon--nucleus scattering experiments and therefore dominantly probe the \textit{charge} density of the nucleus.
Since $^{208}$Pb is a ``doubly-magic'' nucleus (both the number of protons, 82, and number of neutrons, 126, are arranged in fully closed energy shells), it is rather immune to shape deformations, and hence its charge density is well described by a 2pF distribution, with $R$ and $a$ determined to within $1$ and $2$\%, respectively. 
Traditionally, the values for $^{207}$Pb from \cite{DeJager:1987qc} have been used instead of those for $^{208}$Pb from \cite{DeJager:1974liz} when modeling $^{208}$Pb in \MCG\ calculations.~\footnote{There is no clear reason for that, and we speculate that it may simply be an oversight because the $^{208}$Pb parameters were only collected in the earliest, but not in the later \ADND\ publications. 
In any case, the two sets of parameters are essentially the same, as can be seen in \Tab{tab:awR}.}

\begin{figure}[t!]
\centering
\includegraphics[width=0.45\textwidth]{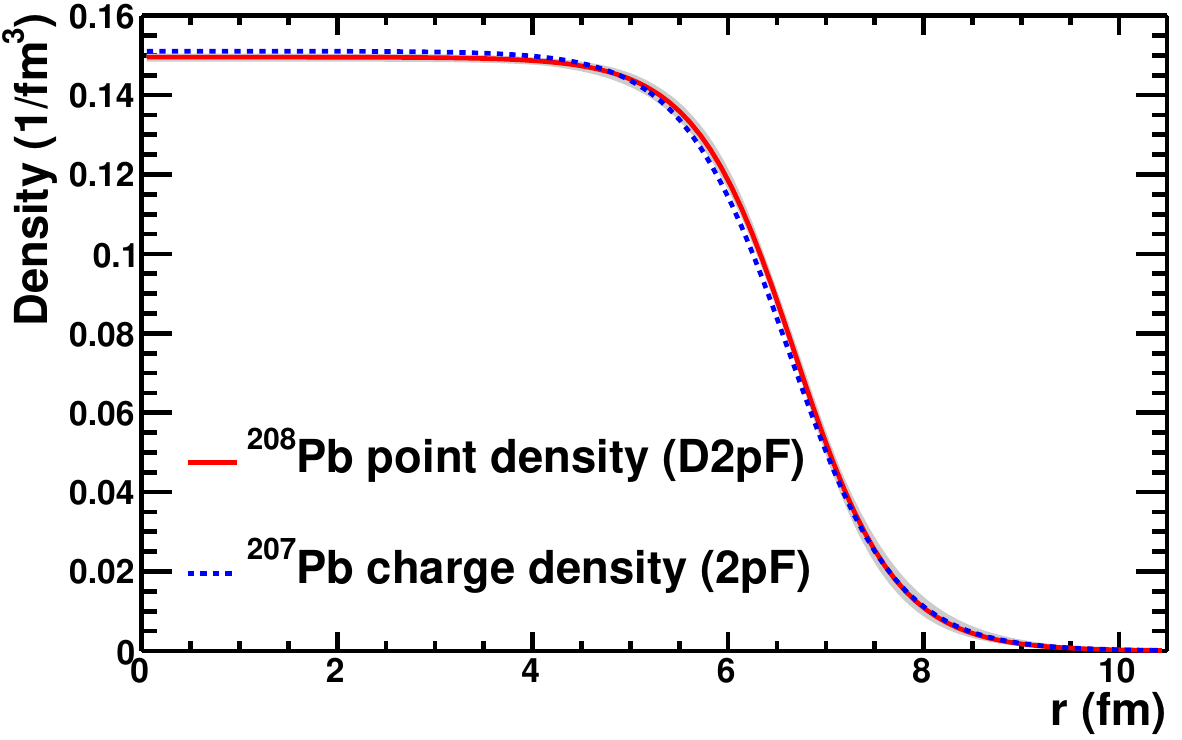} %from macros/mc/plotdens.C
\caption{Nuclear density of $^{208}$Pb for the charge distribution~(2pF) as well as the sum of the proton and neutron point density distributions~(D2pF), corresponding to the parameters listed in \Tab{tab:awR}. 
         The grey band indicates the $1\sigma$ uncertainty for the D2pF distribution.}
\label{fig:2pFs}
\end{figure}
\begin{table}[t!]
\begin{tabular}{l|l|c|c}\hline\hline
\centering
 Charge density                                & Name   & $R$~(fm)            &  $a$~(fm) \\
 \hline
 ${}^{207}$Pb \cite{DeJager:1987qc}             & Pb     &  $6.620 \pm 0.060$  &  $0.546 \pm 0.010$ \\
 ${}^{208}$Pb \cite{DeJager:1974liz}            & Pb*    &  $6.624 \pm 0.035$  &  $0.549 \pm 0.008$ \\
 \hline
 Point density\\      
 \hline
 ${}^{208}$Pb                                   & Pbpn   &    ~                &    ~               \\
 ~~~~proton  \cite{Klos:2007is}                &        &  $6.68 \pm 0.02$    &  $0.447 \pm 0.01 $ \\
 ~~~~neutron \cite{Tarbert:2013jze,Klos:2007is}&        &  $6.69 \pm 0.03$    &  $0.560 \pm 0.03 $ \\
\hline\hline
\end{tabular}
\caption{\label{tab:awR} Nuclear density parameters of Pb for charge and point density distributions. For the neutron point density, the values are averaged as explained in the text. The name of the corresponding profile in the TGlauberMC implementation~\cite{glaucode} is also listed.}
\end{table}

However, since the \MCG\ uses the charge density to place the central locations of each nucleon, a preferred representation is the \textit{point} density distribution, which parameterizes the 2pF function for the centers of the nucleons.  
Transforming from the charge to point distribution involves parameter unfolding which is performed using the proton root-mean-square (\rms) charge radius $\sqrt{\langle r^{2}\rangle} = 0.875$~fm~\cite{Mohr:2015ccw} via the prescription given in \cite{Klos:2007is,Warda:2010qa}.  
The point density 2pF parameter values are slightly smaller than the charge density ones due to the proton's finite spatial extension. 
The corresponding uncertainties on the proton radius density $R_{\rm p}$ have become smaller over the years and are now below $0.5$\%~\cite{Fricke:1995zz,Angeli:2004kvy,Angeli:2013epw}.  
However, the diffusivity parameter for protons, $a_{\rm p}$, is no longer quoted in the more recent \ADND\ tables.  
Moreover, it has been shown that at very large radii (distances greater than $\approx R + 3a$) the 2pF parameterization begins to fail as the measured density falls off faster than a Woods-Saxon distribution~\cite{Wycech:2007jb}.  
This observation can be modeled by letting the diffusivity parameter shrink with increasing $r$, and while the authors provide $a(r)$ for $^{40}$Ca and $^{48}$Ca, they do not provide it for $^{208}$Pb.  
Thus, we maintain the constant-$a$ 2pF form and, in turn, sustain the traditional relatively large uncertainty on the diffusivity parameter of about $2$\% ($\pm 0.01$~fm).

Using the nucleon point density distribution leads to a more realistic placement of the nucleons.  
However, there is evidence that the proton and neutron distributions may not be exactly the  same at the surface of heavy stable nuclei~\cite{barrett1977nuclear}. 
This effect is particularly important in neutron-rich nuclei, such as $^{208}$Pb with a neutron excess of $N/Z\approx 1.5$.  
Protons near the center of the nucleus feel electrostatic repulsion from all directions resulting in an electrostatic equilibrium and a constant charge density.   
However, at $r\!\gtrsim\!6$~fm, where the nucleon density begins to drop, the outermost protons need additional ``skin'' or ``halo'' neutrons in the periphery to counteract the outward Coulomb repulsion and maintain a sufficient nuclear surface tension thereby increasing the overall binding energy. 

To extract the 2pF parameters for neutrons, the Crystal Ball collaboration has performed a measurement via coherent pion photoproduction~\cite{Tarbert:2013jze} while the Low Energy Proton Ring~(LEAR) at CERN has investigated antiproton--nucleus interactions coupled with radiochemistry techniques~\cite{Klos:2007is}.  
The former extracts neutron point density parameters of $R_{\rm n}=6.70\pm0.03$~(stat.)~fm and $a_{\rm n}=0.55\pm0.01$~(stat.)$^{+0.02}_{-0.03}$~(syst.)~fm, while the latter reports comparable values of $R_{\rm n}=6.684\pm0.020$~(stat.)~fm and $a_{\rm n}=0.571$~fm.
These data favor the peripheral neutron distribution in the form of a neutron ``halo'' rather than a neutron ``skin'', \ie\ the neutron distribution is slightly broader than the proton distribution because of its larger diffusivity~($a_{\rm n} - a_{\rm p} \approx 0.1$~fm), but has the same half-radius as the proton distribution~($R_{\rm p} \approx R_{\rm n} \approx 6.7$~fm).
For the LEAR measurement no uncertainty was explicitly reported for $a_{\rm n}$ though the central value is consistent with \cite{Tarbert:2013jze} and we assume $\pm0.03$~fm; both use the same proton charge density parameters taken from the \ADND.  
The neutron parameters are then averaged and listed in \Tab{tab:awR} together with the proton point 2pF parameters. 
The combined point density distribution for proton and neutrons is then the weighted sum of the individual 2pF distributions, which we simulate in the \MCG\ by drawing 82 protons from the proton point 2pF and 128 neutrons from the neutron point 2pF.  
The D2pF distribution is displayed in \Fig{fig:2pFs} with its corresponding $1\sigma$ uncertainty and compared to the traditionally-used charge density distribution.

\begin{figure}[t]
  \centering
  \includegraphics[width=0.45\textwidth]{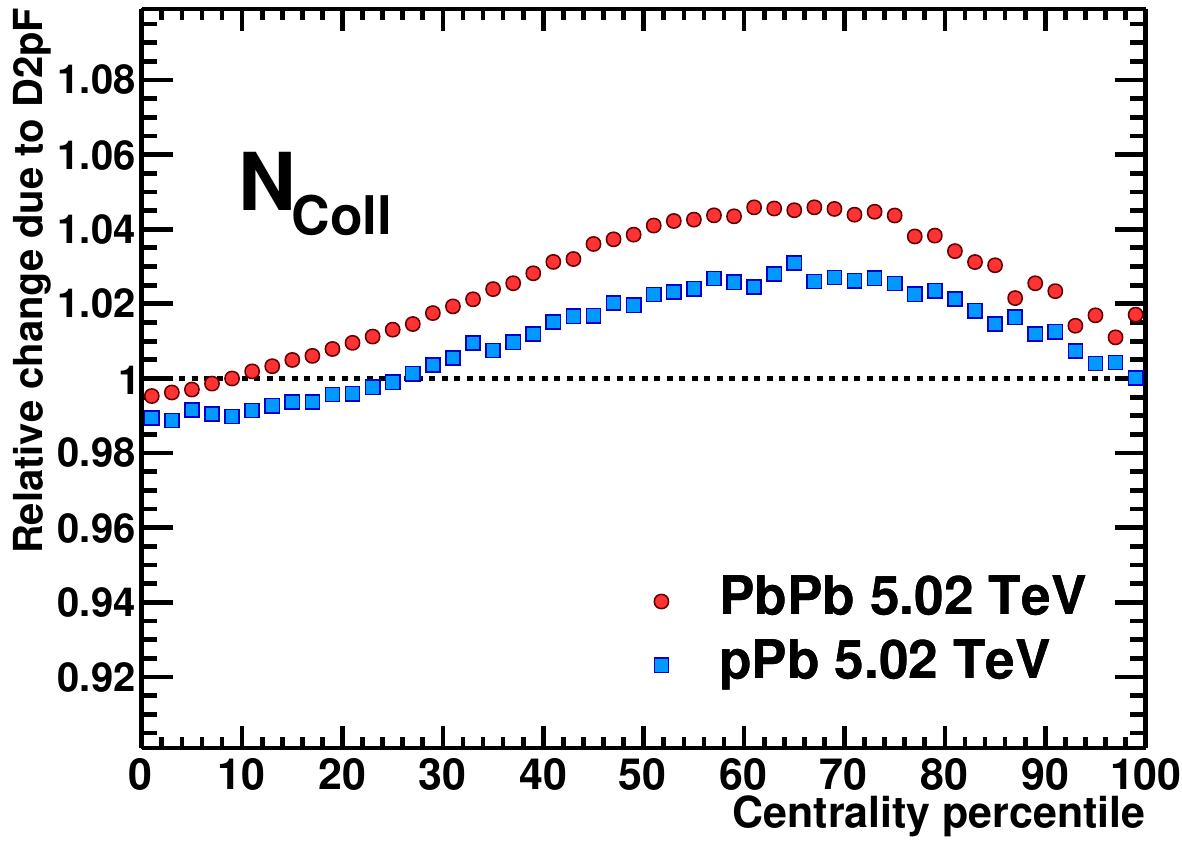} % from macros/tb/DrawNCollPanels_cl.C
  \caption{Relative change in \Ncoll\ in \PbPb\ and \pPb\ collisions at $\snn=5.02$~TeV due to the updated nuclear density profile. 
           The baseline uses the standard 2pF charge density, while the new results are obtained with the D2pF density.}
  \label{fig:NCollChangeD2pF}
\end{figure}

The relative change in \Ncoll\ due to switching from the 2pF charge density to the D2pF point density representation~(while everything else is computed in the traditional approach) is illustrated in \Fig{fig:NCollChangeD2pF}.
In mid-peripheral \PbPb\ collisions, the change results in a maximum $\sim4$\% increase in \Ncoll\ and approximately $2$\% for \pPb\ collisions. 
This is largely driven by the increase of the central radius in the D2pF compared to the 2pF parameterization.

%%%%%%%%%%%%%%%%%%%%%%%%%%%%%%%%%%%%%%%%%%%%%%%%%%%%% 
\subsection{Minimum nucleon separation}
\label{sec:mns}
\begin{figure}[t!]
\centering
\includegraphics[width=0.45\textwidth]{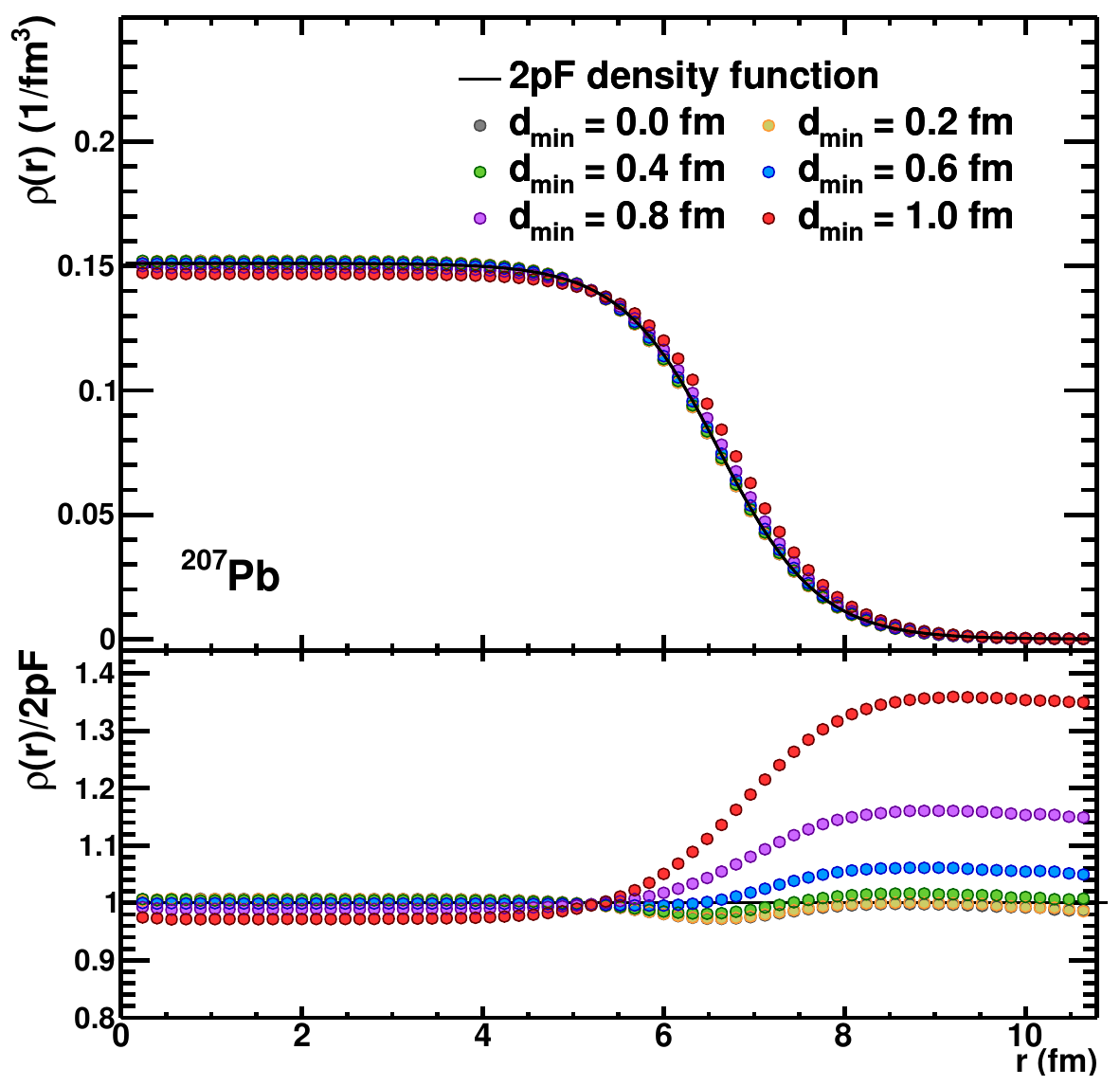} % from macros/density/plot_density_cl.C
  \caption{Nuclear radial density distributions for various values of the inter-nucleon distance $\dmin$ using the ``traditional'' \MCG\ implementation~(top panel) and their ratio to the 2pF profile (bottom panel).
           The deviations from 1 at large $r$ increase with increasing $\dmin$.
}
\label{fig:radialbiasstandard}
\end{figure}

%%%%%%%%%%%%%%%%%%%%%%%%%%%%%%%%%%%%%%%%%%%%%%%%%%%%% 
Prior to this work, varying the inter-nucleon separation from the default value ($0.4$~fm) to its assumed upper limit~($0.8$~fm) led to uncertainties of about $2$\% in the derived Glauber quantities. 
Such a result is somewhat surprising given that, if uniform spherical packing is na\"{\i}vely assumed for nucleons near the center of the nucleus, the typical distance between any two nucleons should be $1.5$--$2$~fm, significantly larger than $\dmin$, and hence the results should not be dramatically affected when varying the latter.  
Traditional \MCG\ implementations place nucleons by first sampling the 2pF distribution and then checking the $\dmin$ requirement with respect to the already placed nucleons.  
When the $\dmin$ requirement is not satisfied, the algorithm discards that nucleon and resamples the 2pF probability distribution.  
This approach results in an overall bias in the constructed radial distribution that propagates to all computed quantities. 
Figure~\ref{fig:radialbiasstandard} shows the resulting deformation in the radial profile due to this bias, which increases with increasing $\dmin$. %for various values of $\dmin$ from $0$ to $1$~fm.
Nucleons are preferentially pushed to larger radii where there is more physical phase-space to fill.  

\begin{figure}[t!]
\centering
\includegraphics[width=0.45\textwidth]{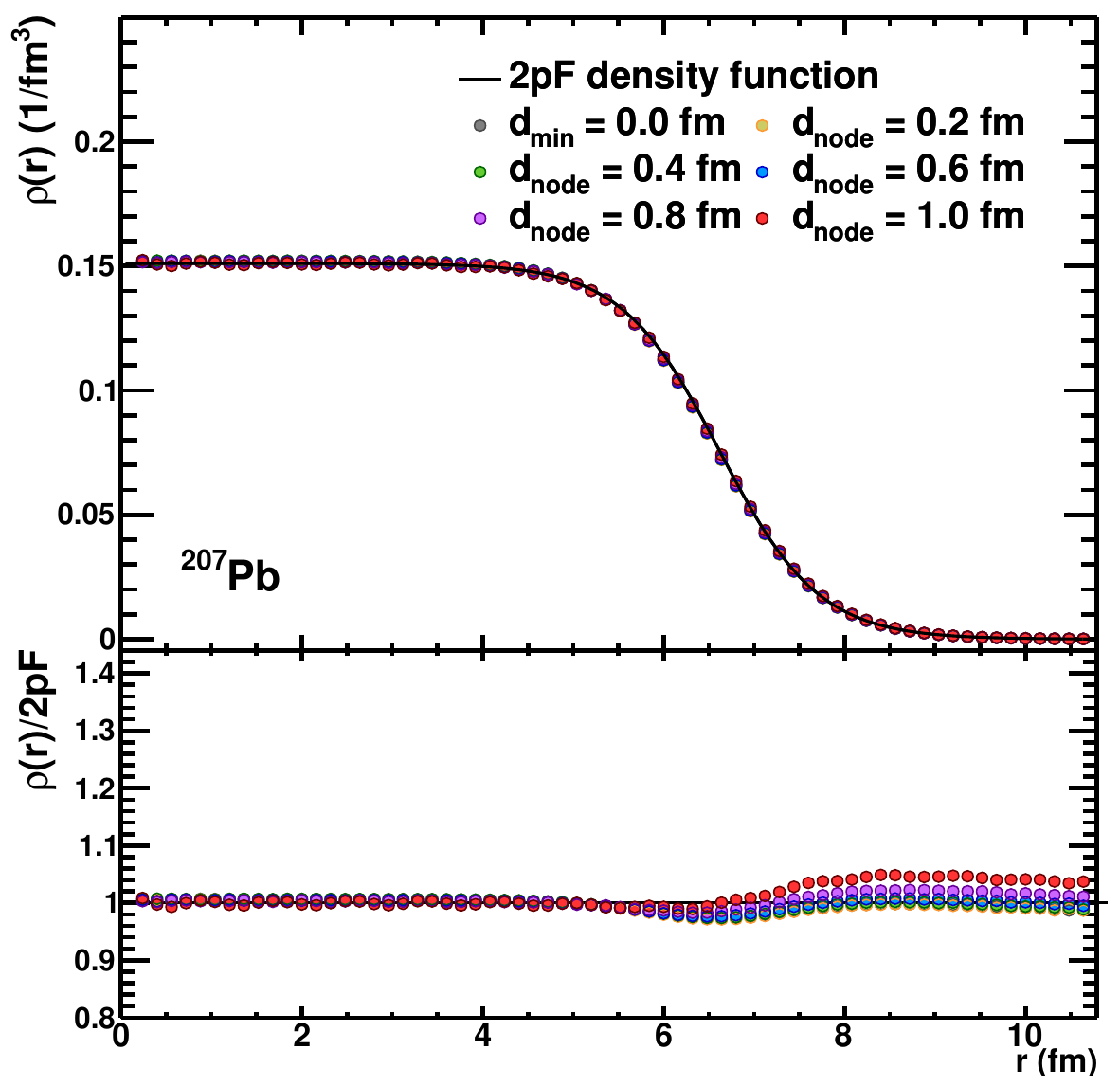} % from macros/density/plot_density_cl.C
  \caption{Nuclear radial density distributions for various values of the inter-nucleon lattice distance $\dnode$~($\equiv \dmin$) using the lattice \MCG\ implementation~(top panel) and their ratio to the 2pF profile (bottom panel).}
\label{fig:radialbiashcp}
\end{figure}

One approach to overcome this effect is to rescale the input profile parameters until the bias brings the resulting density back to the desired 2pF distribution~\cite{Hirano:2009ah,Shou:2014eya}.  
This iterative procedure is cumbersome, unphysical, and not universal for all collisions systems. 
Instead, to remove this bias, we introduce a uniform three-dimensional lattice with a minimum nodal separation~($\dnode$) equivalent to $\dmin$.  
The full physical phase space is sampled by pre-calculating all lattice nodes within a cubic space of $40\times40\times40$~fm$^3$.  
These nodes are sampled uniformly in Cartesian space and subsequently populated with a nucleon according to the 2pF probability distribution. 
Once a node has been populated, it is removed from the sampling.
By apriori restricting the allowable phase-space to exclude overlapping nucleons, the 2pF probability distribution can be sampled without introducing artificial distortions.  
To ensure that regularities in the lattice are avoided, the lattice is randomized event-by-event in azimuthal and polar orientation in addition to being randomly translated laterally in Cartesian space.  
After the implementation of the lattice framework, the density profile remains largely intact and subsequently the centrality variables become stable with respect to $\dmin$ variations.
This is demonstrated in \Fig{fig:radialbiashcp}, which shows the resulting density profiles when varying $\dmin$ by 100\% ($0.4\pm0.4$~fm).  
The results are insensitive to the specific lattice basis used~\cite{hcp}: Hexagonal Close Packing (HCP), Face Centered Cubic, Body Centered Cubic, and Simple Cubic. %were compared.  
Generally, lattices with packing fractions above about $50$\% are indistinguishable for $\dmin<1.2$~fm.  
The HCP lattice was used as the default configuration as it has the most optimal packing fraction of $74$\%.
The insensitivity to the lattice structure is intuitive when considering that less than 0.5\% of nodes inside a radius of about $6.7$~fm are populated when $\dmin = 0.4$~fm.  
As either $\dmin$ is increased to larger than $1.2$~fm or the packing fraction drops significantly below 50\%, the fraction of nodes available will be greatly reduced and distortions start to impact the density distribution.  
It should be noted that, from a technical standpoint, the same result can be achieved with the traditional \MCG\ implementation by discarding the entire nucleus in the event of two nucleons overlapping (rather than only the offending nucleon).  
This, however, is computationally prohibitive and therefore impractical.  

\begin{figure}[t!]
  \centering
  \includegraphics[width=0.45\textwidth]{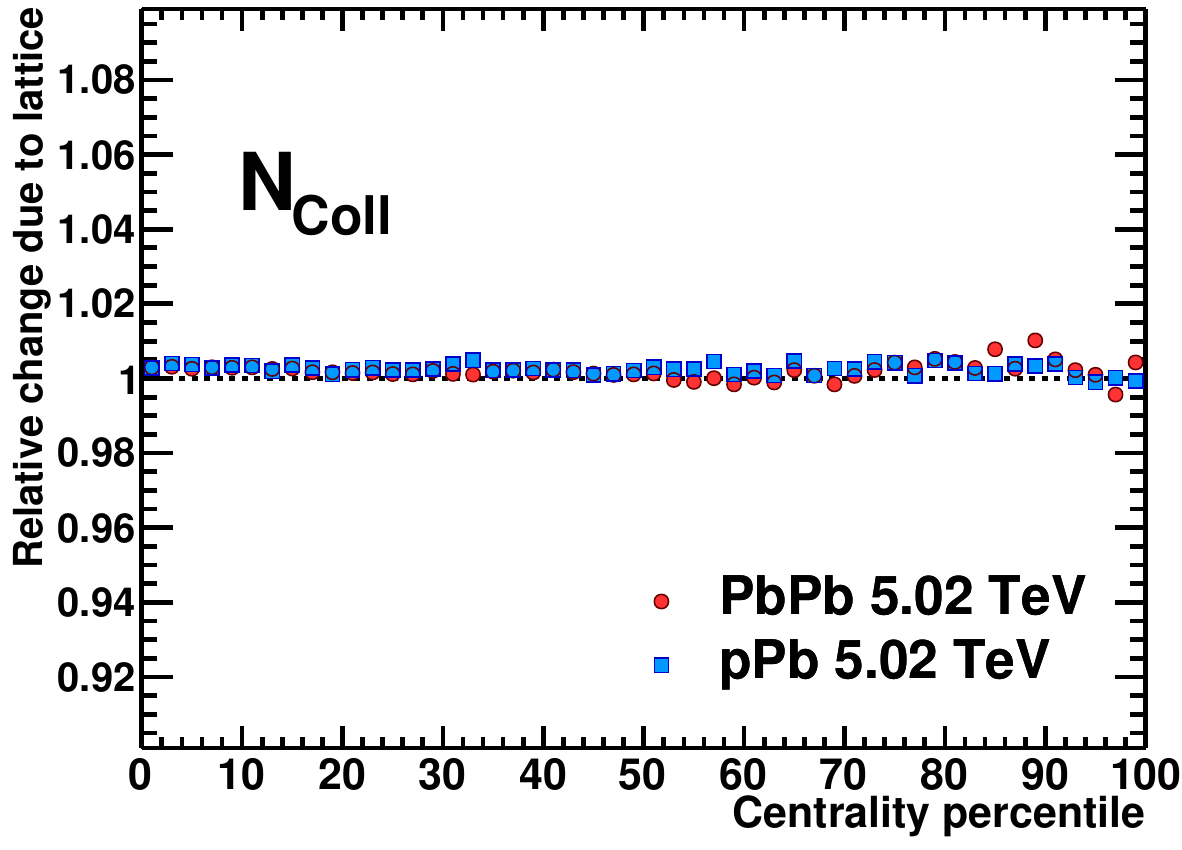} %from from macros/tb/DrawNCollPanels_cl.C
  \caption{Relative change in $\Ncoll$ for \PbPb\ and \pPb\ collisions at $\snn=5.02$~TeV after introducing the nucleon lattice placement algorithm with $\dnode=0.4$~fm. The baseline uses the traditional \MCG\ implementation with $\dmin=0.4$~fm.}
  \label{fig:NCollChangeLattice}
\end{figure}

Figure~\ref{fig:NCollChangeLattice} quantifies the relative change in $\Ncoll$ with respect to the traditional \MCG\ implementation for $\dmin=0.4$~fm for \PbPb\ and \pPb\ collisions. 
The mean value of \Ncoll\ as a function of centrality changes by less than 0.2\%. 
Since the radial profile is not affected by variations of $\dmin$, introducing the lattice to construct the nuclei effectively removes the uncertainty due to the minimum distance between nucleons~(see \Sec{sec:results}).

\begin{figure}[t!]
  \centering
  \includegraphics[width=0.45\textwidth]{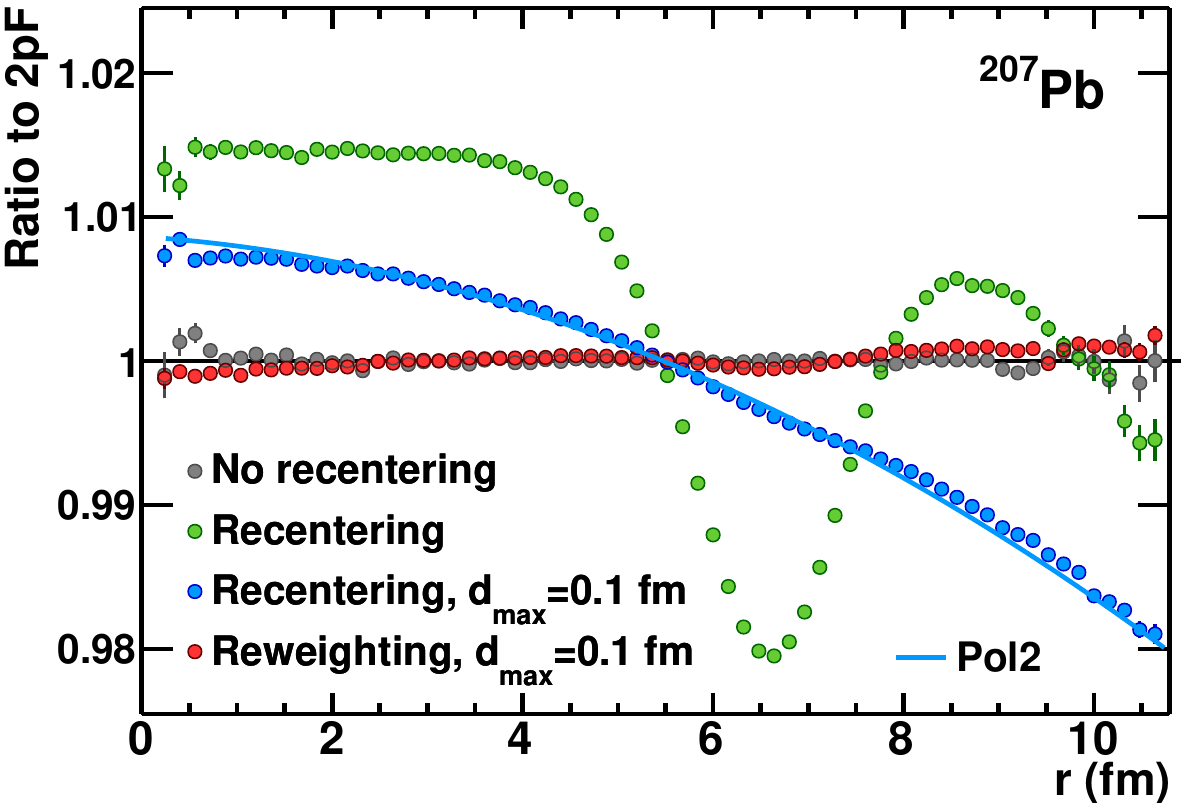} % from macro/density/plot_density_rc.C
  \caption{Ratio of radial density distributions constructed with the standard \MCG\ for $\dmin=0$~fm over that from the 2pF profile for different methods to recenter the nucleons. 
           The \ifarxiv{green }\else{non-monotonous }\fi distribution is obtained with the standard approach including recentering, while the \ifarxiv{grey }\else{uniform }\fi distribution is without recentering.
           The other two cases impose $\dmax=0.1$~fm~(as explained in the text). The \ifarxiv{red }\else{mildly rising }\fi distribution is obtained by dividing~(reweighting) the Pb nucleus radial profile with the 2nd-order polynomial\ifarxiv{~(blue line)}\fi.}
  \label{fig:rhoshift}
\end{figure}

%%%%%%%%%%%%%%%%%%%%%%%%%%%%%%%%%%%%%%%%%%%%%%%%%%%%% 
\subsection{Recentering}
\label{sec:rec}
%%%%%%%%%%%%%%%%%%%%%%%%%%%%%%%%%%%%%%%%%%%%%%%%%%%%% 
Inspecting the bottom panel of \Fig{fig:radialbiashcp} closely, reveals that there are still residual differences of up to a few percent in the radial profile, even when the lattice is used.
Indeed, even for $\dmin=0$~fm, \ie\ without a requirement on the nucleon--nucleon separation, a non-monotonic structure emerges, as can be seen in the ``zoomed-in'' ratio relative to the 2pF profile displayed in \Fig{fig:rhoshift}.
It originates from the recentering algorithm that is usually applied in \MCG\ calculations, since\co{, as demonstrated in \Fig{fig:rhoshift},} without recentering the ratio relative to the 2pF is exactly one.
The traditional \MCG\ approaches~\cite{Alver:2008aq,Loizides:2014vua,Broniowski:2007nz,Rybczynski:2013yba,Loizides:2016djv}, except the HIJING model~\cite{Wang:1991hta}\co{~(and derived initial state models)}, recenter the nucleons by the average of the displacement after having distributed them individually according to the nuclear density profile.
This is also the case for the advanced MC calculation of \Ref{Alvioli:2011sk}, which includes realistic nucleon--nucleon correlations~\cite{Alvioli:2009ab}. %private communication
%There is no reason to assume that the nuclear force would lead to ground state of a nucleus where the cms is not at zero. 
The recentering step is applied to ensure that the center-of-mass of the constructed nucleus coincides with that of the nuclear density from which the nucleon positions were stochastically determined.
Shifting the nucleons by the average displacement, however, introduces a distortion of the radial profile, which increases with decreasing degrees of freedom. 
This effect has been recently discussed in Glauber approaches accounting for subnucleonic degrees of freedom, where the distortion is particularly large when only three partons (quarks) are distributed inside a proton~\cite{Mitchell:2016jio}. 
For a $^{208}$Pb nucleus, the width of the center-of-mass shift is about $0.2$~fm in each direction, and the effect of the associated distortions have so far been ignored.
In order to ensure that the center-of-mass of the constructed nucleus is at $(0,0,0)$, one can only accept constructed nuclei where the average displacement in each direction is small, \eg\ smaller than $\dmax=0.1$~fm.
This requirement leads to more (less) dense radial distributions than the 2pF profile for small~(large) radii, as can be seen in the corresponding ratio~(blue curve) in \Fig{fig:rhoshift}.
The corresponding ratio can be empirically described by a second-order polynomial as $f(r)=1.00863-0.00045r-0.00021r^2$.
Reweighting the original radial profile with $f$, \ie\ using $\rho/f$ to distribute nucleons in the radial direction, allows to correct for the residual bias.
The ratio of the resulting radial distribution relative to the 2pF deviates by less than $0.2$\% from unity.

\begin{figure}[t!]
  \centering
  \includegraphics[width=0.45\textwidth]{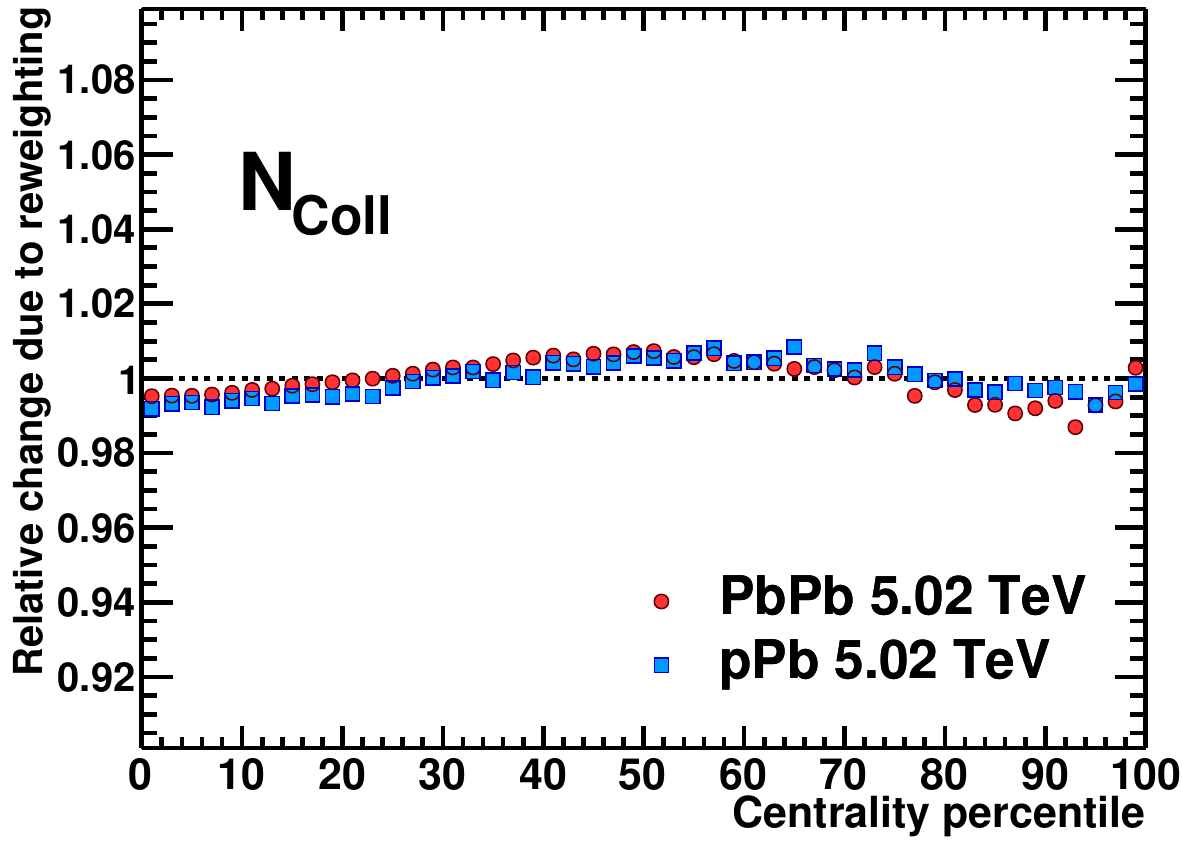} %from macros/tb/DrawNCollPanels_cl.C
  \caption{Relative change in \Ncoll\ in \PbPb\ and \pPb\ collisions at $\snn=5.02$~TeV using the reweighted profile for $\dmin=0.1$~fm and $\dmin=0.4$. 
           The baseline uses the traditional \MCG\ implementation with $\dmin=0.4$~fm.}
  \label{fig:NCollChangeRW}
\end{figure}
\begin{figure}[t!]
  \centering
  \includegraphics[width=0.45\textwidth]{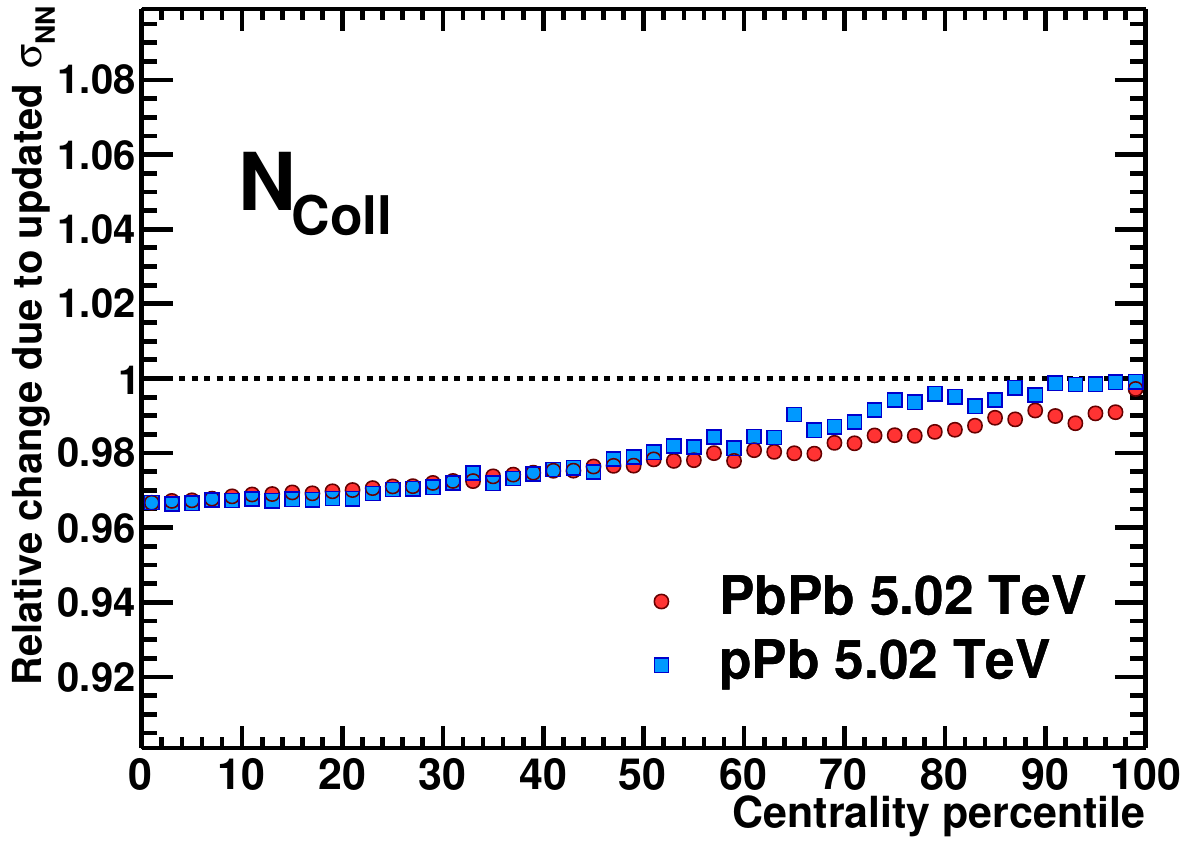} %from macros/tb/DrawNCollPanels_cl.C
  \caption{Relative change in \Ncoll\ in \PbPb\ and \pPb\ collisions at $\snn=5.02$~TeV due to the updated nucleon-nucleon cross section. The baseline uses $\sigmaNN=70$~mb, while the updated value is $67.6$~mb.}
  \label{fig:NCollChangeXsec} 
\end{figure}

%\cleardoublepage
\begin{figure}[t!]
  \centering
  \includegraphics[width=0.5\textwidth]{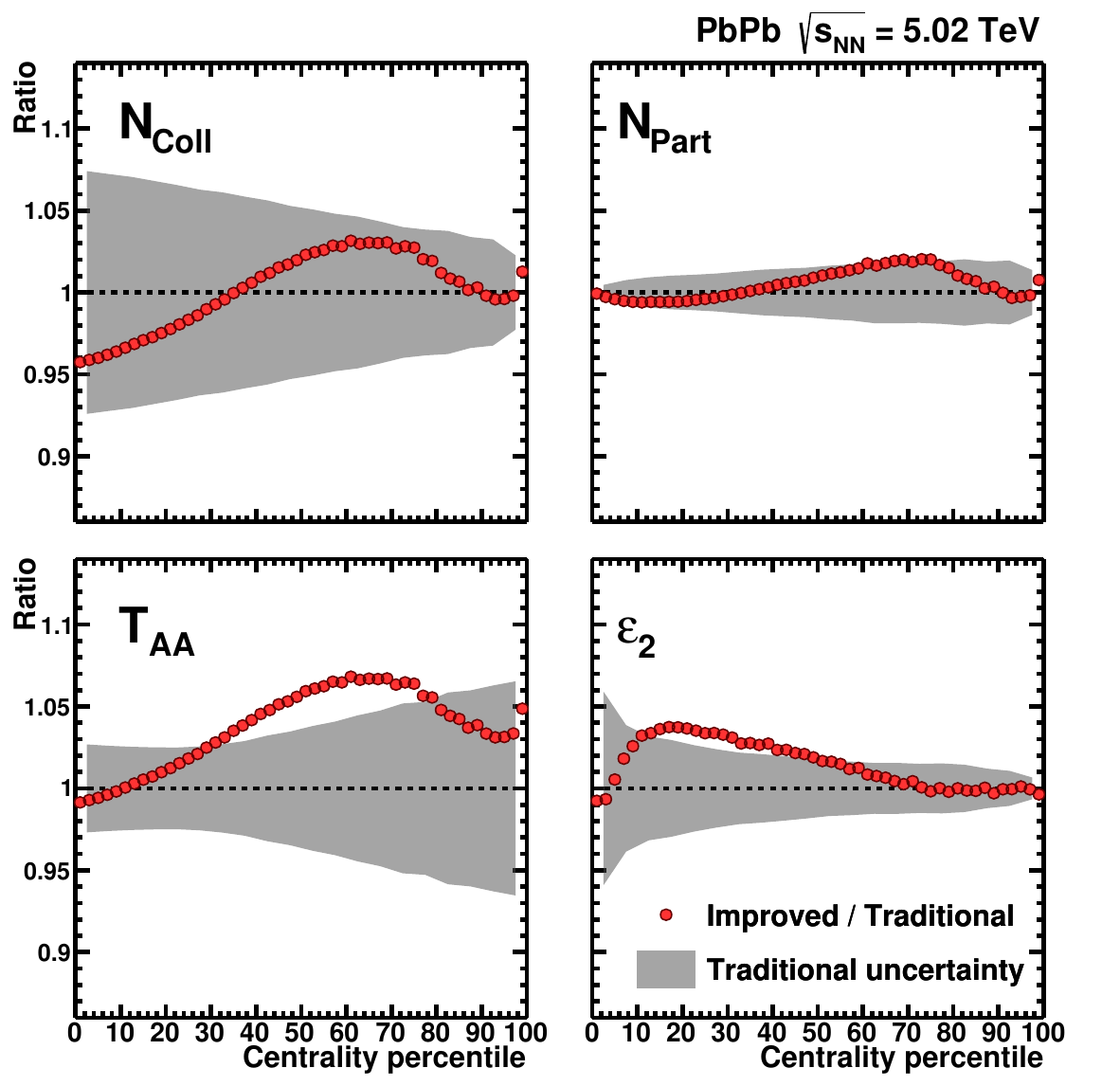} %from macros/tb/DrawErrorPanels.C
  \caption{Ratio of \Ncoll, \Npart, \TAA\ and $\varepsilon_2$ as a function of centrality obtained using the improved approach~(including \snn, lattice, and D2pF changes) over the same quantities obtained with the traditional approach, for \PbPb\ collisions at $\snn=5.02$~TeV. The grey band illustrates the traditionally reported uncertainties.}
  \label{fig:changesNewOldPbPb}
\end{figure}
\begin{figure}[t!]
  \centering
  \includegraphics[width=0.5\textwidth]{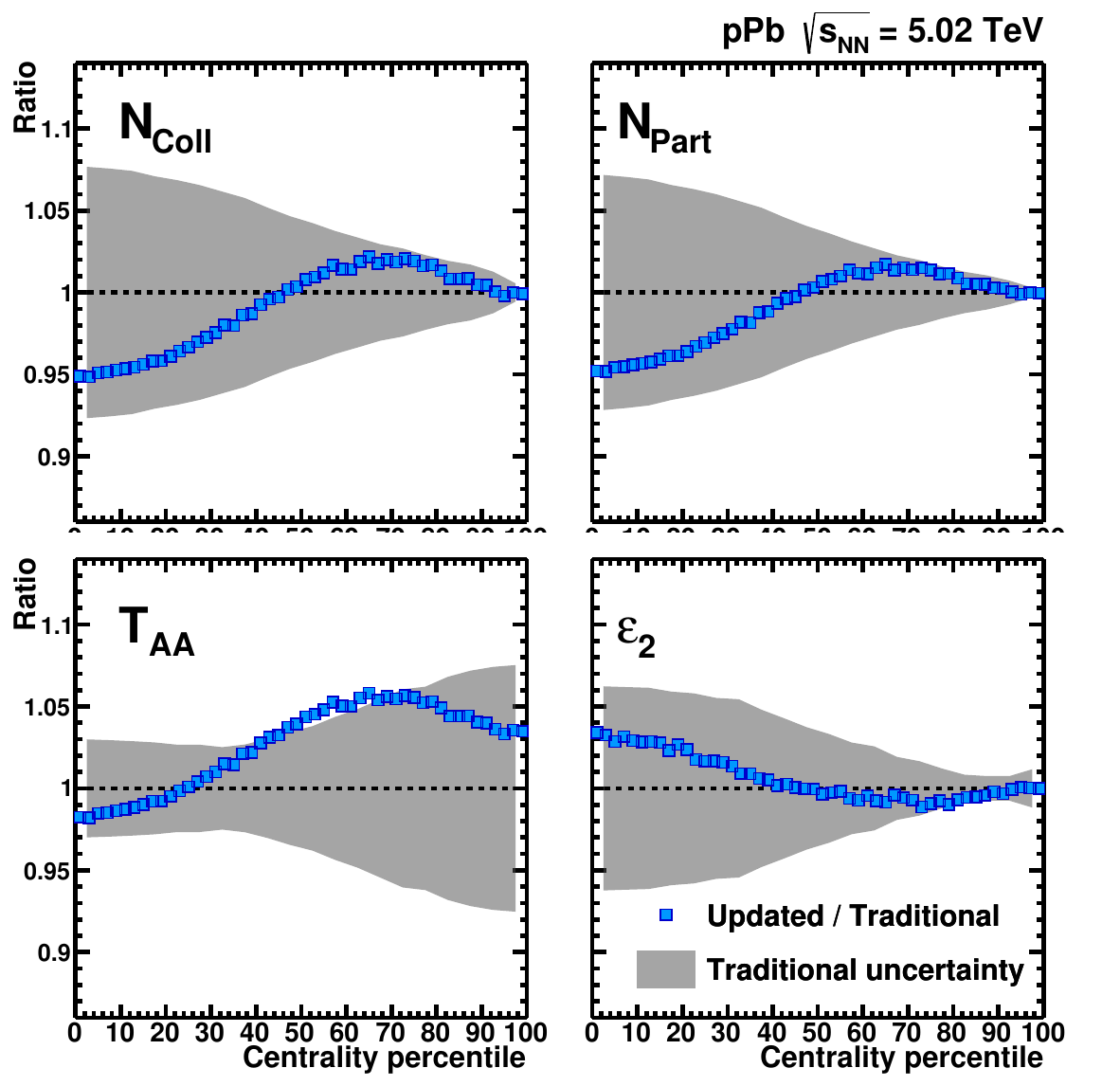} %from macros/tb/DrawErrorPanels.C
  \caption{Ratio of \Ncoll, \Npart, \TAA\ and $\varepsilon_2$  as a function of centrality obtained using the improved approach~(including \snn, lattice, and D2pF changes) over the same quantities obtained with the traditional approach, for \pPb\ collisions at $\snn=5.02$~TeV. The grey band illustrates the traditionally reported uncertainties.}
  \label{fig:changesNewOldpPb}
\end{figure}
\begin{figure}[t!]
  \centering
  \includegraphics[width=0.5\textwidth]{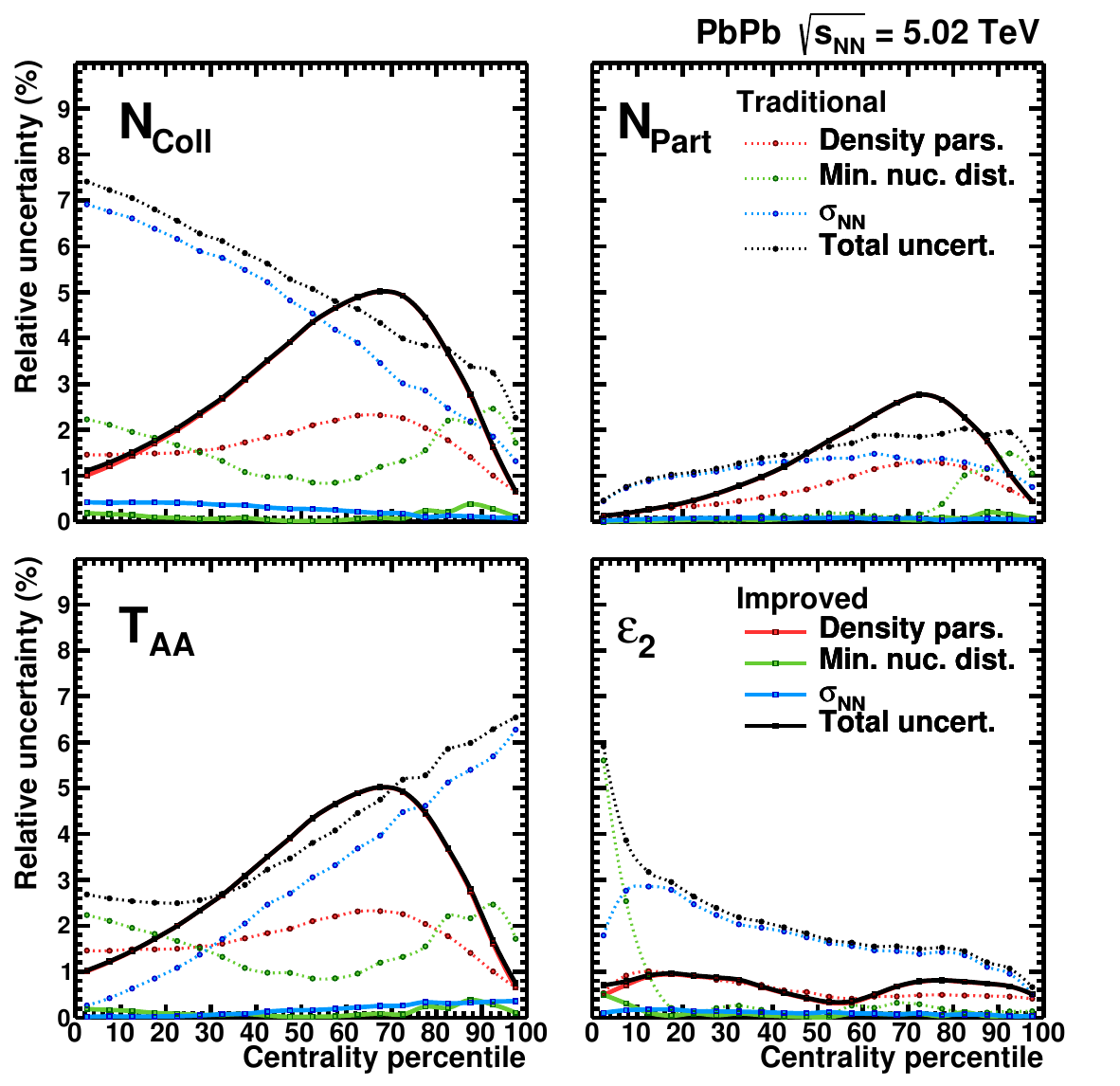} %from macros/tb/DrawErrorPanels.C
  \caption{Relative uncertainties in \Ncoll, \Npart, \TAA\ and $\varepsilon_2$ arising from varying \sigmaNN, as well as inter-nucleon separation and density parameters for the improved and traditional approaches in \PbPb\ collisions at $\snn=5.02$~TeV.}
  \label{fig:CompUncertPbPb}
\end{figure}
\begin{figure}[t!]
  \centering
  \includegraphics[width=0.5\textwidth]{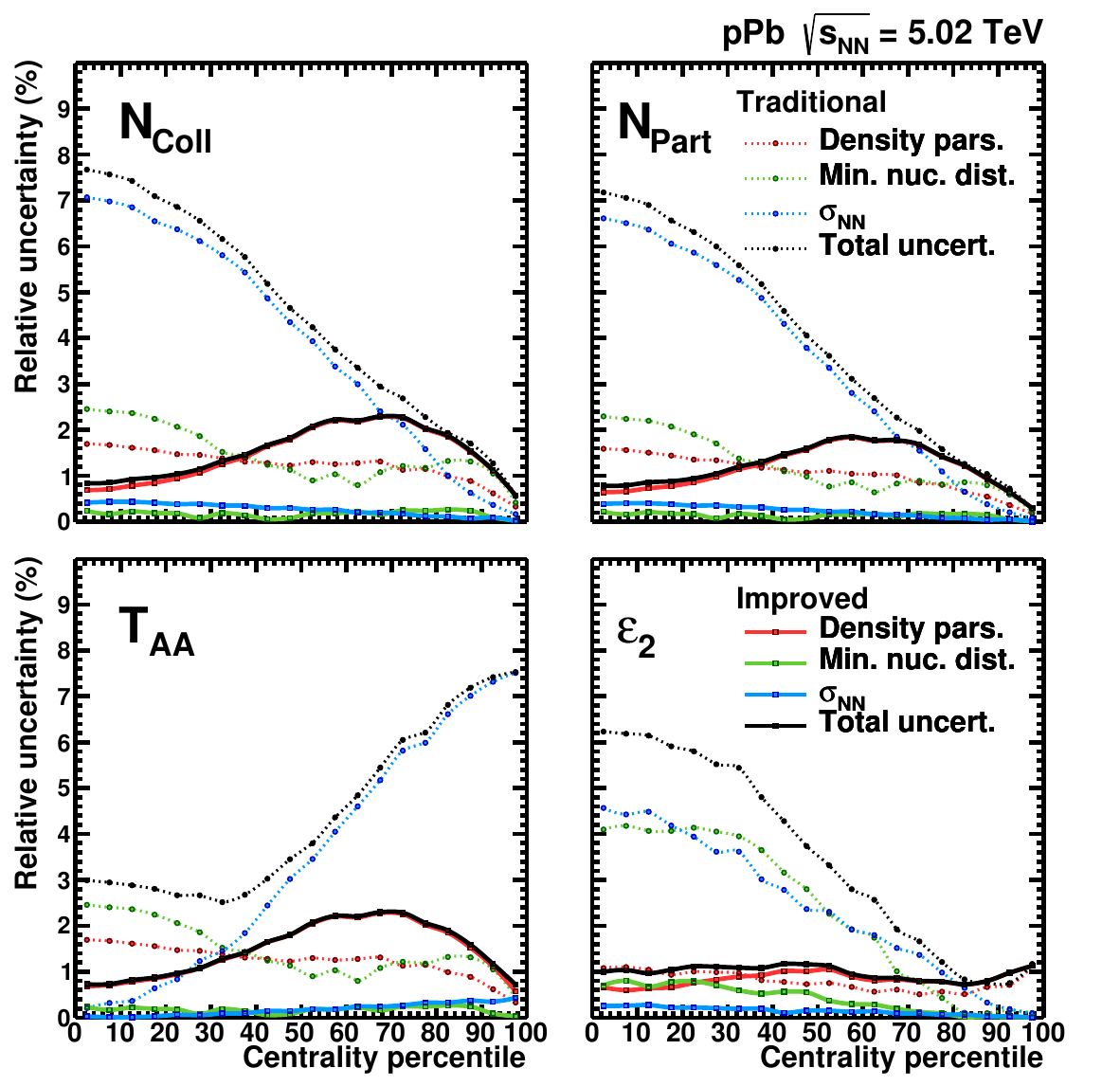} %from macros/tb/DrawErrorPanels.C
  \caption{Relative uncertainties in \Ncoll, \Npart, and \TAA\ and $\varepsilon_2$ arising from varying \sigmaNN\ as well as inter-nucleon separation and density parameters for the improved and traditional approaches in \pPb\ collisions at $\snn=5.02$~TeV.}
  \label{fig:CompUncertpPb}
\end{figure}
\cleardoublepage

The effect on $\Ncoll$ of the reweighted radial profile relative to the standard 2pF profile is quantified in \Fig{fig:NCollChangeRW}, and leads to variations below 1\%.
In particular, for \pPb\ collisions the residual change from recentering to reweighting is much smaller than the modification introduced by recentering alone, which for peripheral collisions is larger than 10\%~(see \Fig{fig:NCollChangeNorec} in the Appendix).

%%%%%%%%%%%%%%%%%%%%%%%%%%%%%%%%%%%%%%%%%%%%%%%%%%%%% 
\subsection{Nucleon--nucleon collision modeling}
\label{sec:nnmod}
%%%%%%%%%%%%%%%%%%%%%%%%%%%%%%%%%%%%%%%%%%%%%%%%%%%%% 
Given that the nucleon-nucleon interaction probability depends on the condition given by \Eq{eq:mc_collisions}, the improved $\sqrt{s}$-parameterization and uncertainty of \sigmaNN\ discussed in Section~\ref{sec:sigmaNN} leads to Glauber quantities that are both more accurate and more precise than before.
To demonstrate the effect of this change for $\snn=5.02$~TeV, the relative change of \Ncoll\ with respect to the previously used \sigmaNN\ value is shown in \Fig{fig:NCollChangeXsec}.
The baseline uses the value of $\sigmaNN=70$~mb, commonly used at the LHC, while the updated value is $67.6$~mb.
As expected, the change is largest for central collisions, namely equal to the ratio of $67.6/70\approx0.97$, while for the most peripheral collisions there is no observable numerical change.

%%%%%%%%%%%%%%%%%%%%%%%%%%%%%%%%%%%%%%%%%%%%%%%%%%%%% 
\section{Results}
\label{sec:results}
%%%%%%%%%%%%%%%%%%%%%%%%%%%%%%%%%%%%%%%%%%%%%%%%%%%%% 
The improvements considered here, including the $\sqrt{s}$-parameterization of $\sigmaNN$, the use of the D2pF profile, plus lattice regularization as well as the recentering and reweighting approach, comprise the {\it improved} \MCG\ approach~\cite{glaucode}, whose parameters are summarized in \Tab{tab:settings}.
To illustrate the differences with the traditional approach, we compare the values of $\Ncoll$, $\Npart$, $\TAA$ and $\varepsilon_2$ computed with both approaches for \PbPb\ and \pPb\ collisions at $\snn=5.02$~TeV in \Figs{fig:changesNewOldPbPb}{fig:changesNewOldpPb}.
The uncertainties due to the 2pF and D2pF parameters, given in \Tab{tab:awR}, were calculated by running 100k \MCG\ events for 100 parameter set variations.  
Each variation allowed each parameter to take a random value within a Gaussian distribution with a width of the $1\sigma$ uncertainty on each parameter.  
The spread of the resulting values quantified by their standard deviation was used as the reported resulting $1\sigma$ uncertainty due to the 2pF and D2pF parameters.
The uncertainties due to $\sigmaNN$ as well as due to the minimum inter-nuclear separation ($\dmin$ and $\dnode$), given in \Tab{tab:settings}, were obtained by running with nominal settings varying each one of the parameters by $\pm\sigma$ at a time, and assigning half of the difference as the corresponding $1\sigma$ uncertainty. 
To obtain the total uncertainties the individual uncertainties due to density profile, $\sigmaNN$, and inter-nucleon separation were added in quadrature.

\Figures{fig:changesNewOldPbPb}{fig:changesNewOldpPb} quantify the changes as the ratio of $\Ncoll$, $\Npart$, $\TAA$ and $\varepsilon_2$ in $5$\%-wide centrality intervals using the improved and traditional approach for \PbPb\ and \pPb\ collisions at $\snn=5.02$~TeV, respectively. %, and \Figs{fig:CompUncertPbPb}{fig:CompUncertpPb} the respective uncertainties as dotted lines for the traditional and full lines for the improved model.
The ratios are compared to the total uncertainties of the traditional approach to illustrate that the central values of the improved results are generally within the previously assigned uncertainties, which were typically dominated by the large uncertainty on $\sigmaNN$, except in the case of $\TAA$ for peripheral \PbPb\ collisions.
Since in $\TAA$ the quite large uncertainties on $\sigmaNN$ cancel out, this quantity is especially sensitive to other small changes introduced by the improvements. 
Our results indicate that, for the relevant centrality classes, previous experimental results on $\RAA$ would have to be scaled up by up to $3$--$5$\%, however ratios of results taking at $\snn=5.02$ and $\snn=2.76$ would not be affected because $\TAA$ would change similarly in both cases. 
We checked that the lattice and traditional approaches lead to identical results for identical settings.
Hence, since it is less computationally intensive, one can also use $\dmin=0.4$~fm in the traditional way~(\ie\ without the lattice) but ignoring the uncertainties introduced from variation to $\dmin=0$ and $0.8$~fm.

\begin{figure}[t!]
\centering
\includegraphics[width=0.45\textwidth]{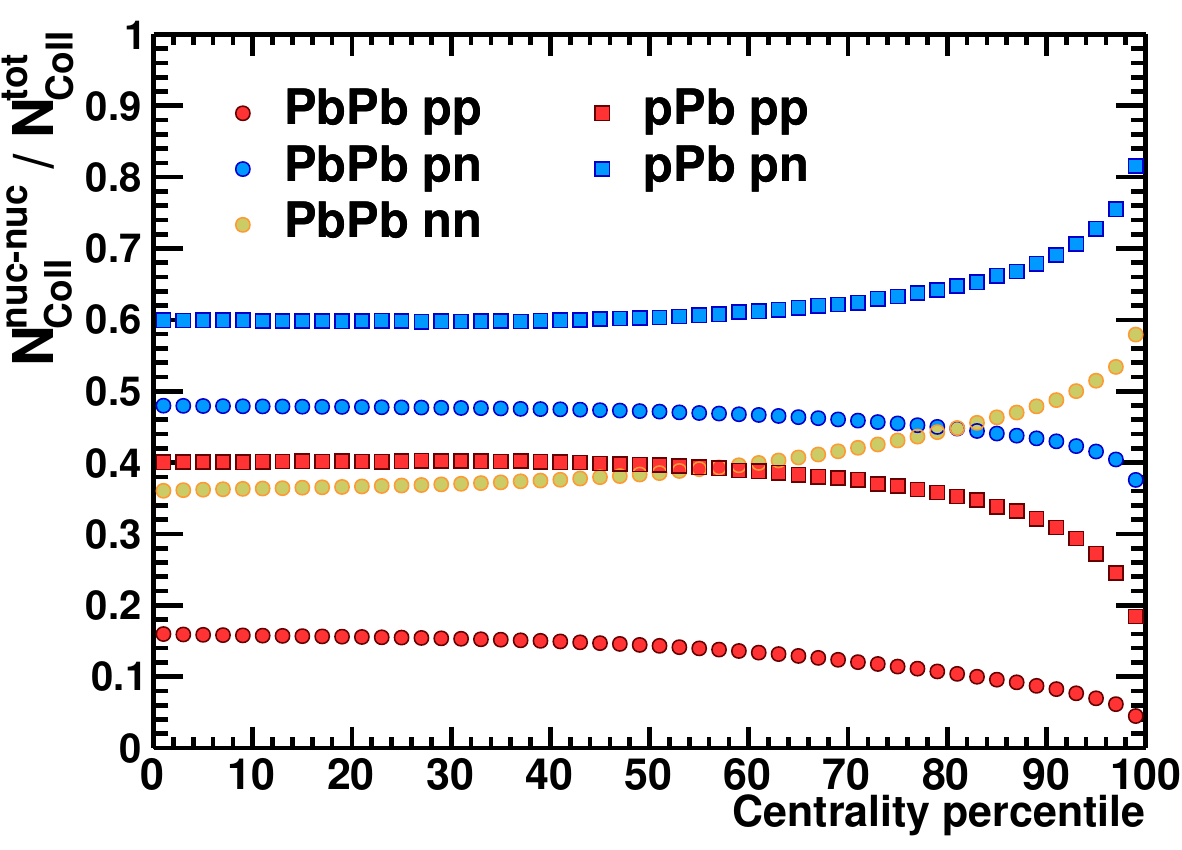} %from macros/tb/DrawNCollTypes.C
  \caption{Average fraction of \pp, \pn, and \nn\ collisions for \PbPb\ and \pPb\ collisions at $\snn=5.02$~TeV  obtained with our improved \MCG\ model.}
\label{fig:Frac_Ncollpn}
\end{figure}
\begin{table}[t!]
\begin{tabular}{c|c|c|c|l}\hline\hline
\centering
 $\snn$ (TeV) & $\sigmaNN$ (mb) & System & $\sigma$ (b)  & Table \\\hline
 2.76           & $61.8\pm0.9$  & \PbPb\ & $7.57\pm0.03$ & \Tab{tab:cent276}\\
 5.02           & $67.6\pm0.6$  & \PbPb\ & $7.66\pm0.03$ & \Tab{tab:cent502}\\
 5.5            & $68.5\pm0.5$  & \PbPb\ & $7.67\pm0.03$ & \Tab{tab:cent55}\\
 10.6           & $75.3\pm0.7$  & \PbPb\ & $7.77\pm0.03$ & \Tab{tab:cent106}\\
 39             & $90.5\pm3.3$  & \PbPb\ & $7.90\pm0.03$ & \Tab{tab:cent39}\\\hline
 5.02           & $67.6\pm0.6$  & \pPb\  & $2.08\pm0.01$ & \Tab{tab:cppb502}\\
 8.16           & $72.5\pm0.5$  & \pPb\  & $2.12\pm0.01$ & \Tab{tab:cppb816}\\
 8.8            & $73.3\pm0.6$  & \pPb\  & $2.13\pm0.01$ & \Tab{tab:cppb88}\\
 17             & $80.6\pm1.5$  & \pPb\  & $2.18\pm0.01$ & \Tab{tab:cppb17}\\
 63             & $96.5\pm4.6$  & \pPb\  & $2.28\pm0.01$ & \Tab{tab:cppb63}\\\hline
\ifnodef
 5.44           & $68.4\pm0.5$  & \XeXe\ & $5.67\pm0.02$ & \Tab{tab:cxexe544} \\
 0.2            & $41.6\pm0.6$  & \AuAu\ & $6.80\pm0.03$ & \Tab{tab:cauau02}\\
 0.2            & $41.6\pm0.6$  & \CuCu\ & $3.43\pm0.03$ & \Tab{tab:ccucu02}\\\hline\hline
\else
 5.44           & $68.4\pm0.5$  & \XeXe\ & $5.67\pm0.02$ & \Tab{tab:cxexe544d} \\
 0.2            & $41.6\pm0.6$  & \AuAu\ & $6.80\pm0.03$ & \Tab{tab:cauau02d}\\
 0.2            & $41.6\pm0.6$  & \CuCu\ & $3.43\pm0.03$ & \Tab{tab:ccucu02d}\\\hline\hline
\fi
\end{tabular} 
\caption{\label{tab:ssum} Values for total \PbPb\ and \pPb\ cross sections~(with statistical uncertainties) at collision energies relevant for the LHC and FCC. 
         For completeness, results for \XeXe\ at $\snn=5.44$~TeV, as well as as \AuAu\ and \CuCu\ collisions at $\snn=0.2$~TeV are also included.
         The values for $\sigmaNN$ are from \Tab{tab:signnvalues}.
         For every collision system the corresponding centrality dependent Glauber quantities can be found in the specified table provided in \Appendix{app:tab}.}
\end{table}

\begin{figure}[t!]
  \centering
\includegraphics[width=0.5\textwidth]{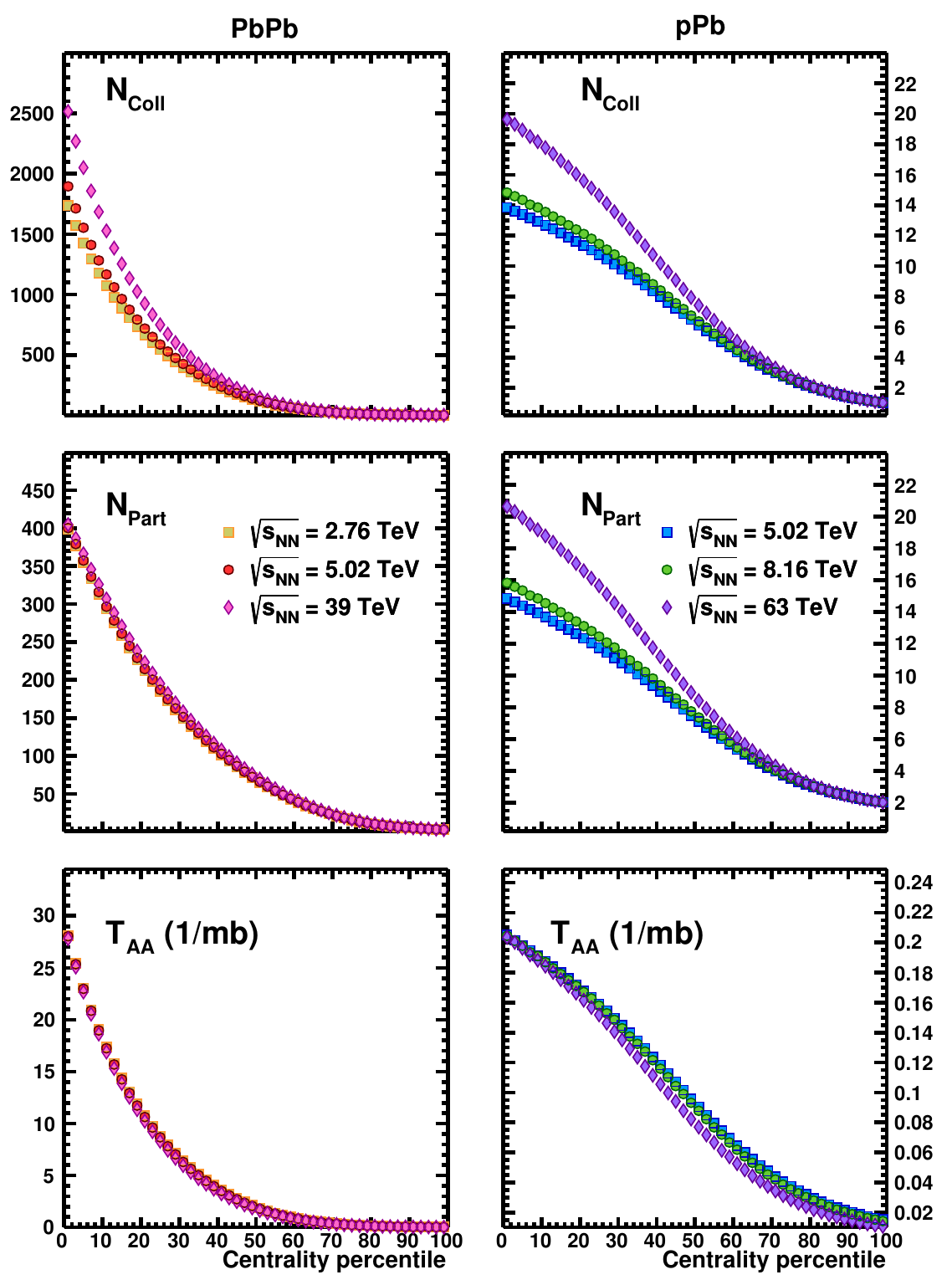} %from macros/tb/DrawResults.C
  \caption{Number of binary collisions (\Ncoll), number of participants (\Npart), and overlap function (\TAA) as a function of centrality for \PbPb\ collisions at $\snn=2.76$, $5.02$ and $39$~TeV~(left columns) and for \pPb\ collisions at $\snn=5.02$, $8.16$ and $63$~TeV~(right columns) using the improved \MCG.}
\label{fig:NcollNpartTaa}
\end{figure}
\begin{figure}[t!]
\centering
\includegraphics[width=0.5\textwidth]{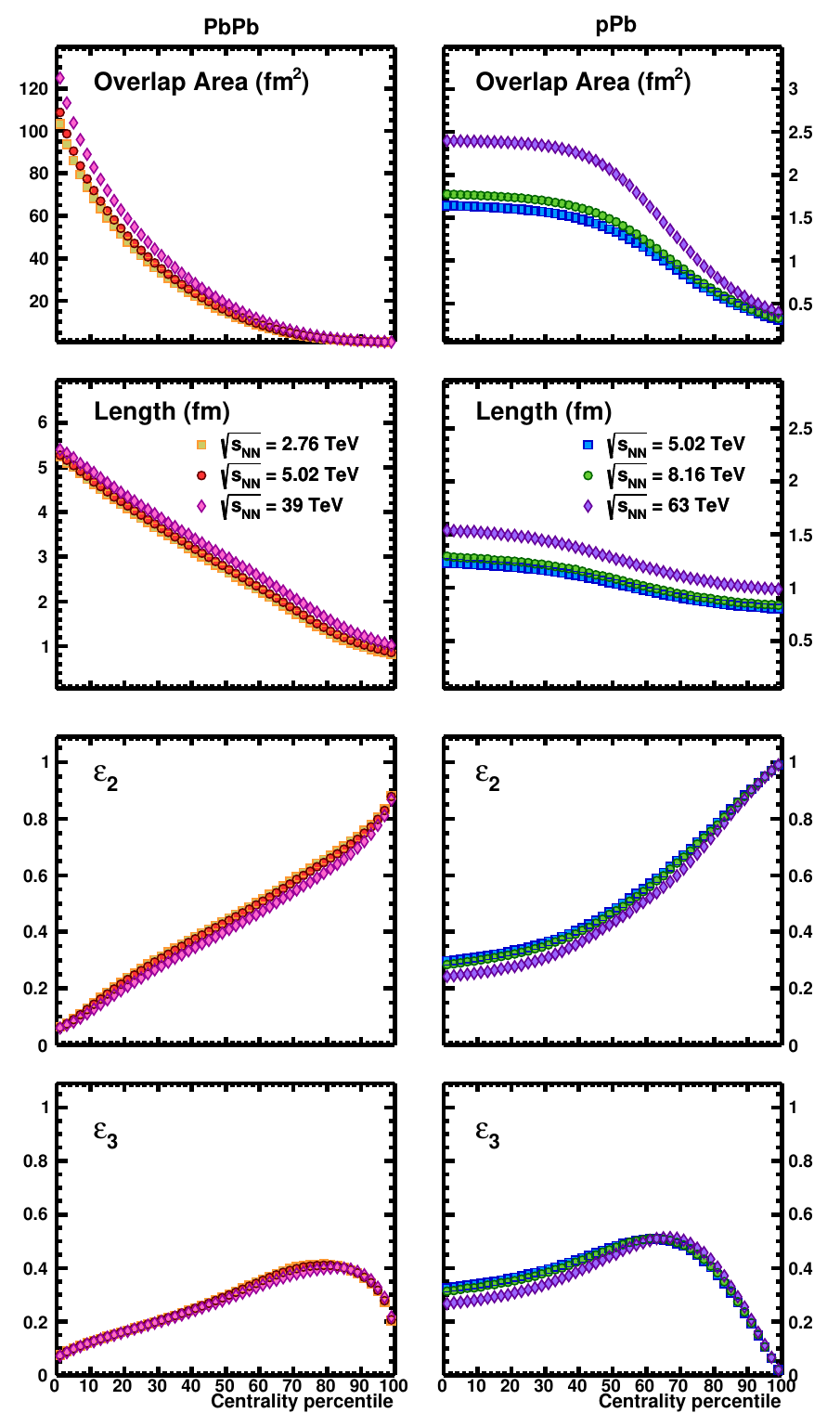} %from macros/tb/DrawResults.C
  \caption{Overlap area ($\AT$), average path length ($L$), participant eccentricity~($\varepsilon_2$) and triangularity~($\varepsilon_3$) as a function of centrality for \PbPb\ collisions at $\snn=2.76$, $5.02$ and $39$~TeV~(left columns) and for \pPb\ collisions at $\snn=5.02$, $8.16$ and $63$~TeV~(right columns) using the improved \MCG.}
\label{fig:AreaLengthEcc}
\end{figure}

\Figures{fig:CompUncertPbPb}{fig:CompUncertpPb} show the respective individual and total uncertainties as dotted lines for the traditional and full lines for the improved model.
The 2pF uncertainties and the minimum inter-nucleon separation ($\dmin$) reach up to about $2$\%, while the (previously) large uncertainty on \sigmaNN\ propagated into up to $7$\% on $\Ncoll$ for central collisions and $\TAA$ for peripheral collisions, and typically dominated the final uncertainty .
In contrast, the uncertainties due to the minimum separation enforced between nucleons by the lattice as well as due to the more precise parameterization of \sigmaNN($\snn$) are quite small, and, in particular, for the inter-nucleon separation nearly negligible in the improved approach.  
This is particularly apparent in the case of $\varepsilon_2$, where the uncertainty related to $\dmin$ was substantial in central collisions for the traditional approach.
The uncertainty due to the D2pF parameters, however, have grown.  
Since the uncertainty on the neutron diffusivity has actually increased to about $5$\% and there are about $50$\% more neutrons than protons in $^{208}$Pb, the inclusion of the D2pF nuclear density description results in a more accurate, though less precise, determination of Glauber quantities.  
For this reason, coupled with the fact that traditional 2pF forms represented the charge density rather than the point density, previously quoted uncertainties based on the 2pF parameters were slightly underestimated. 
For \pPb\ collisions, other experimental uncertainties become dominant, such as those resulting from the event activity class used to determine the centrality~\cite{Adam:2014qja}.
Furthermore, it is important to realize that the spread of the computed quantities in each centrality class is rather large, in particular for peripheral collisions, where the ratios of the standard deviation over the mean of each Glauber quantity can reach up to 80\% (as can be seen in the Tables of \Appendix{app:tab}).

\Figure{fig:Frac_Ncollpn} shows the average fraction of \pp, \pn, and \nn\ collisions for \PbPb\ and \pPb\ collisions at $\snn=5.02$~TeV from the D2pF calculation.  
In peripheral collisions, the \pn\ and \nn\ interactions become more probable due to the extended neutron ``halo'' or ``skin'', and therefore are particularly relevant for precision measurements involving isospin-- (or electric charge--) dependent observables, such as electroweak boson production, in nuclear collisions~\cite{Paukkunen:2015bwa,De:2016ggl,Helenius:2016dsk}.

Finally, we present the number of binary collisions~(\Ncoll), number of participants~(\Npart), and overlap~(\TAA) in \Fig{fig:NcollNpartTaa}, as well as the overlap area~($\AT$), average path length~($L$), participant eccentricity~($\varepsilon_2$), and triangularity~($\varepsilon_3$) in \Fig{fig:AreaLengthEcc} as a function of centrality for \PbPb\ collisions at $\snn=2.76$, $5.02$ and $39$ TeV (left plots), and in \pPb\ collisions at $\snn=5.02$, $8.16$ and $63$~TeV (right plots) using the improved \MCG. 
The inelastic cross sections for all collision systems computed with the improved approach are given in \Tab{tab:ssum}.
\Appendix{app:tab} provides detailed tables for the corresponding quantities in $5$\%-wide centrality classes.

%%%%%%%%%%%%%%%%%%%%%%%%%%%%%%%%%%%%%%%%%%%%%%%%%%%%% 
\section{Summary}
\label{sec:summary}
%%%%%%%%%%%%%%%%%%%%%%%%%%%%%%%%%%%%%%%%%%%%%%%%%%%%% 
We have presented the results of an improved Monte Carlo Glauber model for the calculation of quantities of relevance for collisions involving nuclei at center-of-mass energies of BNL RHIC ($\snn=0.2$~TeV), CERN LHC~($\snn=2.76$--$8.8$~TeV), and proposed future hadron colliders~($\snn\approx 10$--$63$~TeV). 
The corresponding values for the inelastic \pp\ cross section are obtained from a data-driven parametrization resulting in a tenfold reduction of the uncertainties due to the many available measurements at LHC collision energies~(\Fig{fig:sigmappvs}).
We describe the nuclear transverse density with two independent 2-parameter Fermi distributions for protons and neutrons to account for their different densities close to the nuclear periphery. 
Furthermore, we model the nucleon degrees of freedom inside a nucleus using a lattice with a minimum nodal separation to enforce the exclusion of overlapping nucleons without distorting the nuclear density.
Residual small distortions in the generated nuclear densities, resulting from adjusting the nucleon center-of-mass with that of the nucleus, are overcome by appropriately reweighting the original nuclear density.
We demonstrate for collisions at $\snn=5.02$~TeV that the central values of the first four quantities change due to the inclusion of the separated proton and neutron transverse distributions, though they remain typically within the previously assigned systematic uncertainties, while their new associated uncertainties are generally smaller than for earlier calculations~(Figs.~\ref{fig:changesNewOldPbPb}--\ref{fig:CompUncertpPb}).
The number of participant nucleons, binary nucleon--nucleon collisions, nuclear overlap function, participant eccentricity and triangularity, overlap area and average path length are presented in intervals of percentile centrality for lead--lead~(\PbPb) and proton--lead~(\pPb) collisions at all collisions energies~(\Figs{fig:NcollNpartTaa}{fig:AreaLengthEcc}).
Tables for all quantities versus centrality at present and foreseen collision energies involving Pb-nuclei, but also for \XeXe\ at $\snn=5.44$~TeV, and for \AuAu\ and \CuCu\ collisions at $\snn=0.2$~TeV, are provided~(see~\Tab{tab:ssum}).
The source code for the Monte Carlo Glauber program is made publicly available in~\Ref{glaucode}. 
The authors welcome comments on the code and suggestions on how to make it more useful to both experimentalists and theorists.

%%%%%%%%%%%%%%%%%%%%%%%%%%%%%%%%%%%%%%%%%%%%%%%%%%%%% 
\section*{Acknowledgments}
%%%%%%%%%%%%%%%%%%%%%%%%%%%%%%%%%%%%%%%%%%%%%%%%%%%%% 
We are grateful to Klaus Reygers for common work on a first study of the impact of separated neutron and proton densities on the Glauber quantities.
We thank Dan Watts for fruitful discussions regarding \Ref{Tarbert:2013jze}. 
J.K.\ and C.L.\ are supported by the U.S. Department of Energy, Office of Science, Office of Nuclear Physics, under contract numbers DE-FG02-94ER40865 and DE-AC05-00OR22725, respectively.
%%%%%%%%%%%%%%%%%%%%%%%%%%%%%%%%%%%%%%%%%%%%%%%%%%%%% 
\bibliography{biblio} 

\providecommand{\href}[2]{#2}\begingroup\raggedright\begin{thebibliography}{10}

\bibitem{Miller:2007ri}
M.~L. Miller, K.~Reygers, S.~J. Sanders, and P.~Steinberg, ``{Glauber modeling
  in high energy nuclear collisions},''
  \href{http://dx.doi.org/10.1146/annurev.nucl.57.090506.123020}{{\em Ann. Rev.
  Nucl. Part. Sci.} {\bfseries 57} (2007) 205--243},
\href{http://arxiv.org/abs/nucl-ex/0701025}{{\ttfamily arXiv:nucl-ex/0701025
  [nucl-ex]}}.
%%CITATION = NUCL-EX/0701025;%%.

\bibitem{DeJager:1974liz}
C.~W. De~Jager, H.~De~Vries, and C.~De~Vries, ``{Nuclear charge and
  magnetization density distribution parameters from elastic electron
  scattering},''
\href{http://dx.doi.org/10.1016/S0092-640X(74)80002-1}{{\em Atom. Data Nucl.
  Data Tabl.} {\bfseries 14} (1974) 479--508}.
%%CITATION = ADNDA,14,479;%%.

\bibitem{DeJager:1987qc}
H.~De~Vries, C.~W. De~Jager, and C.~De~Vries, ``{Nuclear charge and
  magnetization density distribution parameters from elastic electron
  scattering},''
\href{http://dx.doi.org/10.1016/0092-640X(87)90013-1}{{\em Atom. Data Nucl.
  Data Tabl.} {\bfseries 36} (1987) 495--536}.
%%CITATION = ADNDA,36,495;%%.

\bibitem{Glauber:1970jm}
R.~J. Glauber and G.~Matthiae, ``{High-energy scattering of protons by
  nuclei},''
\href{http://dx.doi.org/10.1016/0550-3213(70)90511-0}{{\em Nucl. Phys.}
  {\bfseries B21} (1970) 135--157}.
%%CITATION = NUPHA,B21,135;%%.

\bibitem{Wang:1991hta}
X.-N. Wang and M.~Gyulassy, ``{HIJING: A Monte Carlo model for multiple jet
  production in \pp, \pA\ and \AA\ collisions},''
\href{http://dx.doi.org/10.1103/PhysRevD.44.3501}{{\em Phys. Rev.} {\bfseries
  D44} (1991) 3501--3516}.
%%CITATION = PHRVA,D44,3501;%%.

\bibitem{Alver:2008aq}
B.~Alver, M.~Baker, C.~Loizides, and P.~Steinberg, ``{The PHOBOS Glauber Monte
  Carlo},''
\href{http://arxiv.org/abs/0805.4411}{{\ttfamily arXiv:0805.4411 [nucl-ex]}}.
%%CITATION = ARXIV:0805.4411;%%.

\bibitem{Loizides:2014vua}
C.~Loizides, J.~Nagle, and P.~Steinberg, ``{Improved version of the PHOBOS
  Glauber Monte Carlo},''
  \href{http://dx.doi.org/10.1016/j.softx.2015.05.001}{{\em SoftwareX}
  {\bfseries 1-2} (2015) 13--18},
\href{http://arxiv.org/abs/1408.2549}{{\ttfamily arXiv:1408.2549 [nucl-ex]}}.
%%CITATION = ARXIV:1408.2549;%%.

\bibitem{Broniowski:2007nz}
W.~Broniowski, M.~Rybczynski, and P.~Bozek, ``{GLISSANDO: Glauber initial-state
  simulation and more...},''
  \href{http://dx.doi.org/10.1016/j.cpc.2008.07.016}{{\em Comput. Phys.
  Commun.} {\bfseries 180} (2009) 69--83},
\href{http://arxiv.org/abs/0710.5731}{{\ttfamily arXiv:0710.5731 [nucl-th]}}.
%%CITATION = ARXIV:0710.5731;%%.

\bibitem{Rybczynski:2013yba}
M.~Rybczynski, G.~Stefanek, W.~Broniowski, and P.~Bozek, ``{GLISSANDO 2:
  GLauber Initial-State Simulation AND mOre..., ver. 2},''
  \href{http://dx.doi.org/10.1016/j.cpc.2014.02.016}{{\em Comput. Phys.
  Commun.} {\bfseries 185} (2014) 1759--1772},
\href{http://arxiv.org/abs/1310.5475}{{\ttfamily arXiv:1310.5475 [nucl-th]}}.
%%CITATION = ARXIV:1310.5475;%%.

\bibitem{Loizides:2016djv}
C.~Loizides, ``{Glauber modeling of high-energy nuclear collisions at the
  subnucleon level},'' \href{http://dx.doi.org/10.1103/PhysRevC.94.024914}{{\em
  Phys. Rev.} {\bfseries C94} no.~2, (2016) 024914},
\href{http://arxiv.org/abs/1603.07375}{{\ttfamily arXiv:1603.07375 [nucl-ex]}}.
%%CITATION = ARXIV:1603.07375;%%.

\bibitem{Klos:2007is}
B.~Klos {\em et~al.}, ``{Neutron density distributions from antiprotonic
  $^{208}$Pb and $^{209}$Bi atoms},''
  \href{http://dx.doi.org/10.1103/PhysRevC.76.014311}{{\em Phys. Rev.}
  {\bfseries C76} (2007) 014311},
\href{http://arxiv.org/abs/nucl-ex/0702016}{{\ttfamily arXiv:nucl-ex/0702016
  [NUCL-EX]}}.
%%CITATION = NUCL-EX/0702016;%%.

\bibitem{Tarbert:2013jze}
C.~M. Tarbert {\em et~al.}, ``{Neutron skin of $^{208}$Pb from coherent pion
  photoproduction},''
  \href{http://dx.doi.org/10.1103/PhysRevLett.112.242502}{{\em Phys. Rev.
  Lett.} {\bfseries 112} no.~24, (2014) 242502},
\href{http://arxiv.org/abs/1311.0168}{{\ttfamily arXiv:1311.0168 [nucl-ex]}}.
%%CITATION = ARXIV:1311.0168;%%.

\bibitem{Paukkunen:2015bwa}
H.~Paukkunen, ``{Neutron skin and centrality classification in high-energy
  heavy-ion collisions at the LHC},''
  \href{http://dx.doi.org/10.1016/j.physletb.2015.04.037}{{\em Phys. Lett.}
  {\bfseries B745} (2015) 73--78},
\href{http://arxiv.org/abs/1503.02448}{{\ttfamily arXiv:1503.02448 [hep-ph]}}.
%%CITATION = ARXIV:1503.02448;%%.

\bibitem{De:2016ggl}
S.~De, ``{The effect of neutron skin on inclusive prompt photon production in
  \PbPb\ collisions at Large Hadron Collider energies},''
  \href{http://dx.doi.org/10.1088/1361-6471/aa5689}{{\em J. Phys.} {\bfseries
  G44} no.~4, (2017) 045104},
\href{http://arxiv.org/abs/1609.09608}{{\ttfamily arXiv:1609.09608 [nucl-th]}}.
%%CITATION = ARXIV:1609.09608;%%.

\bibitem{Helenius:2016dsk}
I.~Helenius, H.~Paukkunen, and K.~J. Eskola, ``{Neutron-skin effect in
  direct-photon and charged hadron-production in \PbPb\ collisions at the
  LHC},'' \href{http://dx.doi.org/10.1140/epjc/s10052-017-4709-9}{{\em Eur.
  Phys. J.} {\bfseries C77} no.~3, (2017) 148},
\href{http://arxiv.org/abs/1606.06910}{{\ttfamily arXiv:1606.06910 [hep-ph]}}.
%%CITATION = ARXIV:1606.06910;%%.

\bibitem{Dainese:2016gch}
A.~Dainese {\em et~al.}, ``{Heavy ions at the Future Circular Collider},''
  \href{http://dx.doi.org/10.23731/CYRM-2017-003.635}{{\em CERN Yellow Report}
  no.~3, (2017) 635--692},
\href{http://arxiv.org/abs/1605.01389}{{\ttfamily arXiv:1605.01389 [hep-ph]}}.
%%CITATION = ARXIV:1605.01389;%%.

\bibitem{glaucode}
``{TGlauberMC on HepForge}.'' \url{http://tglaubermc.hepforge.org/}.
\newblock latest version \version.

\bibitem{Bialas:1976ed}
A.~Bialas, M.~Bleszynski, and W.~Czyz, ``{Multiplicity distributions in
  nucleus-nucleus collisions at high energies},''
\href{http://dx.doi.org/10.1016/0550-3213(76)90329-1}{{\em Nucl. Phys.}
  {\bfseries B111} (1976) 461--476}.
%%CITATION = NUPHA,B111,461;%%.

\bibitem{Kharzeev:1996yx}
D.~Kharzeev, C.~Lourenco, M.~Nardi, and H.~Satz, ``{A Quantitative analysis of
  charmonium suppression in nuclear collisions},''
  \href{http://dx.doi.org/10.1007/s002880050392}{{\em Z. Phys.} {\bfseries C74}
  (1997) 307--318},
\href{http://arxiv.org/abs/hep-ph/9612217}{{\ttfamily arXiv:hep-ph/9612217
  [hep-ph]}}.
%%CITATION = HEP-PH/9612217;%%.

\bibitem{Alver:2006wh}
{\bfseries PHOBOS} Collaboration, B.~Alver {\em et~al.}, ``{System size,
  energy, pseudorapidity, and centrality dependence of elliptic flow},''
  \href{http://dx.doi.org/10.1103/PhysRevLett.98.242302}{{\em Phys. Rev. Lett.}
  {\bfseries 98} (2007) 242302},
\href{http://arxiv.org/abs/nucl-ex/0610037}{{\ttfamily arXiv:nucl-ex/0610037
  [nucl-ex]}}.
%%CITATION = NUCL-EX/0610037;%%.

\bibitem{Alver:2010gr}
B.~Alver and G.~Roland, ``{Collision geometry fluctuations and triangular flow
  in heavy-ion collisions},''
  \href{http://dx.doi.org/10.1103/PhysRevC.82.039903,
  10.1103/PhysRevC.81.054905}{{\em Phys. Rev.} {\bfseries C81} (2010) 054905},
  \href{http://arxiv.org/abs/1003.0194}{{\ttfamily arXiv:1003.0194 [nucl-th]}}.
[Erratum: Phys. Rev.C82,039903(2010)].
%%CITATION = ARXIV:1003.0194;%%.

\bibitem{Teaney:2010vd}
D.~Teaney and L.~Yan, ``{Triangularity and dipole asymmetry in heavy ion
  collisions},'' \href{http://dx.doi.org/10.1103/PhysRevC.83.064904}{{\em Phys.
  Rev.} {\bfseries C83} (2011) 064904},
\href{http://arxiv.org/abs/1010.1876}{{\ttfamily arXiv:1010.1876 [nucl-th]}}.
%%CITATION = ARXIV:1010.1876;%%.

\bibitem{Alver:2008zza}
B.~Alver {\em et~al.}, ``{Importance of correlations and fluctuations on the
  initial source eccentricity in high-energy nucleus-nucleus collisions},''
  \href{http://dx.doi.org/10.1103/PhysRevC.77.014906}{{\em Phys. Rev.}
  {\bfseries C77} (2008) 014906},
\href{http://arxiv.org/abs/0711.3724}{{\ttfamily arXiv:0711.3724 [nucl-ex]}}.
%%CITATION = ARXIV:0711.3724;%%.

\bibitem{Drescher:2007cd}
H.-J. Drescher, A.~Dumitru, C.~Gombeaud, and J.-Y. Ollitrault, ``{The
  Centrality dependence of elliptic flow, the hydrodynamic limit, and the
  viscosity of hot QCD},''
  \href{http://dx.doi.org/10.1103/PhysRevC.76.024905}{{\em Phys. Rev.}
  {\bfseries C76} (2007) 024905},
\href{http://arxiv.org/abs/0704.3553}{{\ttfamily arXiv:0704.3553 [nucl-th]}}.
%%CITATION = ARXIV:0704.3553;%%.

\bibitem{Drees:2003zh}
A.~Drees, H.~Feng, and J.~Jia, ``{Medium induced jet absorption at RHIC},''
  \href{http://dx.doi.org/10.1103/PhysRevC.71.034909}{{\em Phys. Rev.}
  {\bfseries C71} (2005) 034909},
\href{http://arxiv.org/abs/nucl-th/0310044}{{\ttfamily arXiv:nucl-th/0310044
  [nucl-th]}}.
%%CITATION = NUCL-TH/0310044;%%.

\bibitem{Dainese:2004te}
A.~Dainese, C.~Loizides, and G.~Paic, ``{Leading-particle suppression in high
  energy nucleus-nucleus collisions},''
  \href{http://dx.doi.org/10.1140/epjc/s2004-02077-x}{{\em Eur. Phys. J.}
  {\bfseries C38} (2005) 461--474},
\href{http://arxiv.org/abs/hep-ph/0406201}{{\ttfamily arXiv:hep-ph/0406201
  [hep-ph]}}.
%%CITATION = HEP-PH/0406201;%%.

\bibitem{Abelev:2013qoq}
{\bfseries ALICE} Collaboration, B.~Abelev {\em et~al.}, ``{Centrality
  determination of \PbPb\ collisions at $\snn=2.76$ TeV with ALICE},''
  \href{http://dx.doi.org/10.1103/PhysRevC.88.044909}{{\em Phys. Rev.}
  {\bfseries C88} no.~4, (2013) 044909},
\href{http://arxiv.org/abs/1301.4361}{{\ttfamily arXiv:1301.4361 [nucl-ex]}}.
%%CITATION = ARXIV:1301.4361;%%.

\bibitem{Alner:1986iy}
{\bfseries UA5} Collaboration, G.~J. Alner {\em et~al.}, ``{Antiproton-proton
  cross sections at 200 and 900 GeV c.m. energy},''
\href{http://dx.doi.org/10.1007/BF01552491}{{\em Z. Phys.} {\bfseries C32}
  (1986) 153--161}.
%%CITATION = ZEPYA,C32,153;%%.

\bibitem{Amos:1990jh}
{\bfseries E710} Collaboration, N.~A. Amos {\em et~al.}, ``{A luminosity
  independent measurement of the $\bar{\rm p}$p total cross section at
  $\sqrt{s}=1.8$ Tev},''
\href{http://dx.doi.org/10.1016/0370-2693(90)90973-A}{{\em Phys. Lett.}
  {\bfseries B243} (1990) 158--164}.
%%CITATION = PHLTA,B243,158;%%.

\bibitem{Amos:1991bp}
{\bfseries E710} Collaboration, N.~A. Amos {\em et~al.}, ``{Measurement of
  $\rho$, the ratio of the real to imaginary part of the $\bar{\rmp}$p forward
  elastic scattering amplitude, at $\sqrt{s}=1.8$ TeV},''
\href{http://dx.doi.org/10.1103/PhysRevLett.68.2433}{{\em Phys. Rev. Lett.}
  {\bfseries 68} (1992) 2433--2436}.
%%CITATION = PRLTA,68,2433;%%.

\bibitem{Abe:1993xy}
{\bfseries CDF} Collaboration, F.~Abe {\em et~al.}, ``{Measurement of the
  $\bar{p}p$ total cross-section at $\sqrt{s}=546$ GeV and 1800 GeV},''
\href{http://dx.doi.org/10.1103/PhysRevD.50.5550}{{\em Phys. Rev.} {\bfseries
  D50} (1994) 5550--5561}.
%%CITATION = PHRVA,D50,5550;%%.

\bibitem{Abe:1993xx}
{\bfseries CDF} Collaboration, F.~Abe {\em et~al.}, ``{Measurement of small
  angle $\bar{p}p$ elastic scattering at $\sqrt{s}=546$ GeV and 1800 GeV},''
\href{http://dx.doi.org/10.1103/PhysRevD.50.5518}{{\em Phys. Rev.} {\bfseries
  D50} (1994) 5518--5534}.
%%CITATION = PHRVA,D50,5518;%%.

\bibitem{Abelev:2012sea}
{\bfseries ALICE} Collaboration, B.~Abelev {\em et~al.}, ``{Measurement of
  inelastic, single- and double-diffraction cross sections in proton--proton
  collisions at the LHC with ALICE},''
  \href{http://dx.doi.org/10.1140/epjc/s10052-013-2456-0}{{\em Eur. Phys. J.}
  {\bfseries C73} no.~6, (2013) 2456},
\href{http://arxiv.org/abs/1208.4968}{{\ttfamily arXiv:1208.4968 [hep-ex]}}.
%%CITATION = ARXIV:1208.4968;%%.

\bibitem{Aad:2011eu}
{\bfseries ATLAS} Collaboration, G.~Aad {\em et~al.}, ``{Measurement of the
  inelastic proton--proton cross section at $\sqrt{s}=7$ TeV with the ATLAS
  Detector},'' \href{http://dx.doi.org/10.1038/ncomms1472}{{\em Nature Commun.}
  {\bfseries 2} (2011) 463},
\href{http://arxiv.org/abs/1104.0326}{{\ttfamily arXiv:1104.0326 [hep-ex]}}.
%%CITATION = ARXIV:1104.0326;%%.

\bibitem{Aad:2014dca}
{\bfseries ATLAS} Collaboration, G.~Aad {\em et~al.}, ``{Measurement of the
  total cross section from elastic scattering in \pp\ collisions at
  $\sqrt{s}=7$ TeV with the ATLAS detector},''
  \href{http://dx.doi.org/10.1016/j.nuclphysb.2014.10.019}{{\em Nucl. Phys.}
  {\bfseries B889} (2014) 486--548},
\href{http://arxiv.org/abs/1408.5778}{{\ttfamily arXiv:1408.5778 [hep-ex]}}.
%%CITATION = ARXIV:1408.5778;%%.

\bibitem{Aaboud:2016ijx}
{\bfseries ATLAS} Collaboration, M.~Aaboud {\em et~al.}, ``{Measurement of the
  total cross section from elastic scattering in \pp\ collisions at
  $\sqrt{s}=8$ TeV with the ATLAS detector},''
  \href{http://dx.doi.org/10.1016/j.physletb.2016.08.020}{{\em Phys. Lett.}
  {\bfseries B761} (2016) 158--178},
\href{http://arxiv.org/abs/1607.06605}{{\ttfamily arXiv:1607.06605 [hep-ex]}}.
%%CITATION = ARXIV:1607.06605;%%.

\bibitem{Aaboud:2016mmw}
{\bfseries ATLAS} Collaboration, M.~Aaboud {\em et~al.}, ``{Measurement of the
  inelastic proton-proton cross section at $\sqrt{s}=13$ TeV with the ATLAS
  detector at the LHC},''
  \href{http://dx.doi.org/10.1103/PhysRevLett.117.182002}{{\em Phys. Rev.
  Lett.} {\bfseries 117} no.~18, (2016) 182002},
\href{http://arxiv.org/abs/1606.02625}{{\ttfamily arXiv:1606.02625 [hep-ex]}}.
%%CITATION = ARXIV:1606.02625;%%.

\bibitem{Chatrchyan:2012nj}
{\bfseries CMS} Collaboration, S.~Chatrchyan {\em et~al.}, ``{Measurement of
  the inelastic proton-proton cross section at $\sqrt{s}=7$ TeV},''
  \href{http://dx.doi.org/10.1016/j.physletb.2013.03.024}{{\em Phys. Lett.}
  {\bfseries B722} (2013) 5--27},
\href{http://arxiv.org/abs/1210.6718}{{\ttfamily arXiv:1210.6718 [hep-ex]}}.
%%CITATION = ARXIV:1210.6718;%%.

\bibitem{VanHaevermaet:2016gnh}
{\bfseries CMS} Collaboration, H.~Van~Haevermaet, ``{Measurement of the
  inelastic proton--proton cross section at $\sqrt{s}$ = 13 TeV},'' in {\em
  {Proceedings, 24th International Workshop on Deep-Inelastic Scattering and
  Related Subjects (DIS 2016): Hamburg, Germany, April 11-25, 2016}}.
\newblock 2016.
\newblock
\href{http://arxiv.org/abs/1607.02033}{{\ttfamily arXiv:1607.02033 [hep-ex]}}.
\newblock
%%CITATION = ARXIV:1607.02033;%%.

\bibitem{Aaij:2014vfa}
{\bfseries LHCb} Collaboration, R.~Aaij {\em et~al.}, ``{Measurement of the
  inelastic pp cross-section at a centre-of-mass energy of $\sqrt{s}=7$ TeV},''
  \href{http://dx.doi.org/10.1007/JHEP02(2015)129}{{\em JHEP} {\bfseries 02}
  (2015) 129},
\href{http://arxiv.org/abs/1412.2500}{{\ttfamily arXiv:1412.2500 [hep-ex]}}.
%%CITATION = ARXIV:1412.2500;%%.

\bibitem{Antchev:2011vs}
{\bfseries TOTEM} Collaboration, G.~Antchev {\em et~al.}, ``{First measurement
  of the total proton-proton cross section at the LHC energy of $\sqrt{s}=7$
  TeV},'' \href{http://dx.doi.org/10.1209/0295-5075/96/21002}{{\em Europhys.
  Lett.} {\bfseries 96} (2011) 21002},
\href{http://arxiv.org/abs/1110.1395}{{\ttfamily arXiv:1110.1395 [hep-ex]}}.
%%CITATION = ARXIV:1110.1395;%%.

\bibitem{Antchev:2013iaa}
{\bfseries TOTEM} Collaboration, G.~Antchev {\em et~al.},
  ``{Luminosity-independent measurements of total, elastic and inelastic
  cross-sections at $\sqrt{s}=7$ TeV},''
\href{http://dx.doi.org/10.1209/0295-5075/101/21004}{{\em Europhys. Lett.}
  {\bfseries 101} (2013) 21004}.
%%CITATION = EULEE,101,21004;%%.

\bibitem{Antchev:2013paa}
{\bfseries TOTEM} Collaboration, G.~Antchev {\em et~al.},
  ``{Luminosity-independent measurement of the proton-proton total cross
  section at $\sqrt{s}=8$ TeV},''
\href{http://dx.doi.org/10.1103/PhysRevLett.111.012001}{{\em Phys. Rev. Lett.}
  {\bfseries 111} no.~1, (2013) 012001}.
%%CITATION = PRLTA,111,012001;%%.

\bibitem{Collaboration:2012wt}
{\bfseries Pierre Auger} Collaboration, P.~Abreu {\em et~al.}, ``{Measurement
  of the proton-air cross-section at $\sqrt{s}=57$ TeV with the Pierre Auger
  Observatory},'' \href{http://dx.doi.org/10.1103/PhysRevLett.109.062002}{{\em
  Phys. Rev. Lett.} {\bfseries 109} (2012) 062002},
\href{http://arxiv.org/abs/1208.1520}{{\ttfamily arXiv:1208.1520 [hep-ex]}}.
%%CITATION = ARXIV:1208.1520;%%.

\bibitem{dEnterria:2016oxo}
D.~d'Enterria and T.~Pierog, ``{Global properties of proton-proton collisions
  at $\sqrt{s}$ = 100 TeV},''
  \href{http://dx.doi.org/10.1007/JHEP08(2016)170}{{\em JHEP} {\bfseries 08}
  (2016) 170},
\href{http://arxiv.org/abs/1604.08536}{{\ttfamily arXiv:1604.08536 [hep-ph]}}.
%%CITATION = ARXIV:1604.08536;%%.

\bibitem{Froissart:1961ux}
M.~Froissart, ``{Asymptotic behavior and subtractions in the Mandelstam
  representation},''
\href{http://dx.doi.org/10.1103/PhysRev.123.1053}{{\em Phys. Rev.} {\bfseries
  123} (1961) 1053--1057}.
%%CITATION = PHRVA,123,1053;%%.

\bibitem{Cudell:2002xe}
{\bfseries COMPETE} Collaboration, J.~R. Cudell, V.~V. Ezhela, P.~Gauron,
  K.~Kang, {\relax Yu}.~V. Kuyanov, S.~B. Lugovsky, E.~Martynov, B.~Nicolescu,
  E.~A. Razuvaev, and N.~P. Tkachenko, ``{Benchmarks for the forward
  observables at RHIC, the Tevatron Run II and the LHC},''
  \href{http://dx.doi.org/10.1103/PhysRevLett.89.201801}{{\em Phys. Rev. Lett.}
  {\bfseries 89} (2002) 201801},
\href{http://arxiv.org/abs/hep-ph/0206172}{{\ttfamily arXiv:hep-ph/0206172
  [hep-ph]}}.
%%CITATION = HEP-PH/0206172;%%.

\bibitem{ALICE:2012aa}
{\bfseries ALICE} Collaboration, B.~Abelev {\em et~al.}, ``{Measurement of the
  cross section for electromagnetic dissociation with neutron emission in
  \PbPb\ collisions at $\snn=2.76$ TeV},''
  \href{http://dx.doi.org/10.1103/PhysRevLett.109.252302}{{\em Phys. Rev.
  Lett.} {\bfseries 109} (2012) 252302},
\href{http://arxiv.org/abs/1203.2436}{{\ttfamily arXiv:1203.2436 [nucl-ex]}}.
%%CITATION = ARXIV:1203.2436;%%.

\bibitem{Khachatryan:2015zaa}
{\bfseries CMS} Collaboration, V.~Khachatryan {\em et~al.}, ``{Measurement of
  the inelastic cross section in proton–lead collisions at $\sqrt {s_{NN}}=$
  5.02 TeV},'' \href{http://dx.doi.org/10.1016/j.physletb.2016.06.027}{{\em
  Phys. Lett.} {\bfseries B759} (2016) 641--662},
\href{http://arxiv.org/abs/1509.03893}{{\ttfamily arXiv:1509.03893 [hep-ex]}}.
%%CITATION = ARXIV:1509.03893;%%.

\bibitem{Abelev:2014epa}
{\bfseries ALICE} Collaboration, B.~B. Abelev {\em et~al.}, ``{Measurement of
  visible cross sections in proton-lead collisions at $\snn$ = 5.02 TeV in van
  der Meer scans with the ALICE detector},''
  \href{http://dx.doi.org/10.1088/1748-0221/9/11/P11003}{{\em JINST} {\bfseries
  9} no.~11, (2014) P11003},
\href{http://arxiv.org/abs/1405.1849}{{\ttfamily arXiv:1405.1849 [nucl-ex]}}.
%%CITATION = ARXIV:1405.1849;%%.

\bibitem{Adam:2014qja}
{\bfseries ALICE} Collaboration, J.~Adam {\em et~al.}, ``{Centrality dependence
  of particle production in \pPb\ collisions at $\snn=5.02$ TeV},''
  \href{http://dx.doi.org/10.1103/PhysRevC.91.064905}{{\em Phys. Rev.}
  {\bfseries C91} no.~6, (2015) 064905},
\href{http://arxiv.org/abs/1412.6828}{{\ttfamily arXiv:1412.6828 [nucl-ex]}}.
%%CITATION = ARXIV:1412.6828;%%.

\bibitem{Fricke:1995zz}
G.~Fricke, C.~Bernhardt, K.~Heilig, L.~A. Schaller, L.~Schellenberg, E.~B.
  Shera, and C.~W. de~Jager, ``{Nuclear Ground State Charge Radii from
  Electromagnetic Interactions},''
\href{http://dx.doi.org/10.1006/adnd.1995.1007}{{\em Atom. Data Nucl. Data
  Tabl.} {\bfseries 60} (1995) 177}.
%%CITATION = ADNDA,60,177;%%.

\bibitem{Angeli:2004kvy}
I.~Angeli, ``{A consistent set of nuclear rms charge radii: properties of the
  radius surface R(N,Z)},''
\href{http://dx.doi.org/10.1016/j.adt.2004.04.002}{{\em Atom. Data Nucl. Data
  Tabl.} {\bfseries 87} no.~2, (2004) 185--206}.
%%CITATION = ADNDA,87,185;%%.

\bibitem{Angeli:2013epw}
I.~Angeli and K.~P. Marinova, ``{Table of experimental nuclear ground state
  charge radii: An update},''
\href{http://dx.doi.org/10.1016/j.adt.2011.12.006}{{\em Atom. Data Nucl. Data
  Tabl.} {\bfseries 99} no.~1, (2013) 69--95}.
%%CITATION = ADNDA,99,69;%%.

\bibitem{Mohr:2015ccw}
P.~J. Mohr, D.~B. Newell, and B.~N. Taylor, ``{CODATA Recommended values of the
  fundamental physical constants: 2014},''
  \href{http://dx.doi.org/10.1103/RevModPhys.88.035009}{{\em Rev. Mod. Phys.}
  {\bfseries 88} no.~3, (2016) 035009},
\href{http://arxiv.org/abs/1507.07956}{{\ttfamily arXiv:1507.07956
  [physics.atom-ph]}}.
%%CITATION = ARXIV:1507.07956;%%.

\bibitem{Warda:2010qa}
M.~Warda, X.~Vinas, X.~Roca-Maza, and M.~Centelles, ``{Analysis of bulk and
  surface contributions in the neutron skin of nuclei},''
  \href{http://dx.doi.org/10.1103/PhysRevC.81.054309}{{\em Phys. Rev.}
  {\bfseries C81} (2010) 054309},
\href{http://arxiv.org/abs/1003.5225}{{\ttfamily arXiv:1003.5225 [nucl-th]}}.
%%CITATION = ARXIV:1003.5225;%%.

\bibitem{Wycech:2007jb}
S.~Wycech, F.~J. Hartmann, J.~Jastrzebski, B.~Klos, A.~Trzcinska, and T.~von
  Egidy, ``{Nuclear surface studies with antiprotonic atom X-rays},''
  \href{http://dx.doi.org/10.1103/PhysRevC.76.034316}{{\em Phys. Rev.}
  {\bfseries C76} (2007) 034316},
\href{http://arxiv.org/abs/nucl-th/0702029}{{\ttfamily arXiv:nucl-th/0702029
  [NUCL-TH]}}.
%%CITATION = NUCL-TH/0702029;%%.

\bibitem{barrett1977nuclear}
R.~C. Barrett and D.~F. Jackson, {\em Nuclear sizes and structure}.
\newblock Oxford University Press, 1977.

\bibitem{Hirano:2009ah}
T.~Hirano and Y.~Nara, ``{Eccentricity fluctuation effects on elliptic flow in
  relativistic heavy ion collisions},''
  \href{http://dx.doi.org/10.1103/PhysRevC.79.064904}{{\em Phys. Rev.}
  {\bfseries C79} (2009) 064904},
\href{http://arxiv.org/abs/0904.4080}{{\ttfamily arXiv:0904.4080 [nucl-th]}}.
%%CITATION = ARXIV:0904.4080;%%.

\bibitem{Shou:2014eya}
Q.~Y. Shou, Y.~G. Ma, P.~Sorensen, A.~H. Tang, F.~Videbæk, and H.~Wang,
  ``{Parameterization of deformed nuclei for Glauber modeling in Relativistic
  Heavy Ion Collisions},''
  \href{http://dx.doi.org/10.1016/j.physletb.2015.07.078}{{\em Phys. Lett.}
  {\bfseries B749} (2015) 215--220},
\href{http://arxiv.org/abs/1409.8375}{{\ttfamily arXiv:1409.8375 [nucl-th]}}.
%%CITATION = ARXIV:1409.8375;%%.

\bibitem{hcp}
P.~Krishna and D.~Pandey, {\em {Close-packed structures}}.
\newblock {University College Cardiff Press}, 1981.
\newblock Download
  \href{https://www.iucr.org/__data/assets/pdf_file/0015/13254/5.pdf}{here}.

\bibitem{Alvioli:2011sk}
M.~Alvioli, H.~Holopainen, K.~J. Eskola, and M.~Strikman, ``{Initial state
  anisotropies and their uncertainties in ultrarelativistic heavy-ion
  collisions from the Monte Carlo Glauber model},''
  \href{http://dx.doi.org/10.1103/PhysRevC.85.034902}{{\em Phys. Rev.}
  {\bfseries C85} (2012) 034902},
\href{http://arxiv.org/abs/1112.5306}{{\ttfamily arXiv:1112.5306 [hep-ph]}}.
%%CITATION = ARXIV:1112.5306;%%.

\bibitem{Alvioli:2009ab}
M.~Alvioli, H.~J. Drescher, and M.~Strikman, ``{A Monte Carlo generator of
  nucleon configurations in complex nuclei including nucleon--nucleon
  correlations},'' \href{http://dx.doi.org/10.1016/j.physletb.2009.08.067}{{\em
  Phys. Lett.} {\bfseries B680} (2009) 225--230},
\href{http://arxiv.org/abs/0905.2670}{{\ttfamily arXiv:0905.2670 [nucl-th]}}.
%%CITATION = ARXIV:0905.2670;%%.

\bibitem{Mitchell:2016jio}
J.~T. Mitchell, D.~V. Perepelitsa, M.~J. Tannenbaum, and P.~W. Stankus,
  ``{Tests of constituent-quark generation methods which maintain both the
  nucleon center of mass and the desired radial distribution in Monte Carlo
  Glauber models},'' \href{http://dx.doi.org/10.1103/PhysRevC.93.054910}{{\em
  Phys. Rev.} {\bfseries C93} no.~5, (2016) 054910},
\href{http://arxiv.org/abs/1603.08836}{{\ttfamily arXiv:1603.08836 [nucl-ex]}}.
%%CITATION = ARXIV:1603.08836;%%.

\bibitem{overlap}
{Dariusz Miskowiec}, ``{Nuclear Overlap Calculation}.''
  \url{http://web-docs.gsi.de/~misko/overlap/}.
\newblock {Code adapted to match the MC implementation as closely as possible}.

\bibitem{Hofstadter:1956qs}
R.~Hofstadter, ``{Electron scattering and nuclear structure},''
\href{http://dx.doi.org/10.1103/RevModPhys.28.214}{{\em Rev. Mod. Phys.}
  {\bfseries 28} (1956) 214--254}.
%%CITATION = RMPHA,28,214;%%.

\bibitem{Alvioli:2013vk}
M.~Alvioli and M.~Strikman, ``{Color fluctuation effects in proton-nucleus
  collisions},'' \href{http://dx.doi.org/10.1016/j.physletb.2013.04.042}{{\em
  Phys. Lett.} {\bfseries B722} (2013) 347--354},
\href{http://arxiv.org/abs/1301.0728}{{\ttfamily arXiv:1301.0728 [hep-ph]}}.
%%CITATION = ARXIV:1301.0728;%%.

\bibitem{Corke:2011yy}
R.~Corke and T.~Sjostrand, ``{Multiparton Interactions with an x-dependent
  Proton Size},'' \href{http://dx.doi.org/10.1007/JHEP05(2011)009}{{\em JHEP}
  {\bfseries 05} (2011) 009},
\href{http://arxiv.org/abs/1101.5953}{{\ttfamily arXiv:1101.5953 [hep-ph]}}.
%%CITATION = ARXIV:1101.5953;%%.

\bibitem{Rybczynski:2013mla}
M.~Rybczyński and Z.~Wlodarczyk, ``{The nucleon–nucleon collision profile
  and cross section fluctuations},''
  \href{http://dx.doi.org/10.1088/0954-3899/41/1/015106}{{\em J. Phys.}
  {\bfseries G41} (2013) 015106},
\href{http://arxiv.org/abs/1307.0636}{{\ttfamily arXiv:1307.0636 [nucl-th]}}.
%%CITATION = ARXIV:1307.0636;%%.

\bibitem{Jia:2009mq}
J.~Jia, ``{Influence of the nucleon-nucleon collision geometry on the
  determination of the nuclear modification factor for nucleon-nucleus and
  nucleus-nucleus collisions},''
  \href{http://dx.doi.org/10.1016/j.physletb.2009.10.044}{{\em Phys. Lett.}
  {\bfseries B681} (2009) 320--325},
\href{http://arxiv.org/abs/0907.4175}{{\ttfamily arXiv:0907.4175 [nucl-th]}}.
%%CITATION = ARXIV:0907.4175;%%.

\bibitem{Morsch:2017brb}
C.~Loizides and A.~Morsch, ``{Absence of jet quenching in peripheral
  nucleus–nucleus collisions},''
  \href{http://dx.doi.org/10.1016/j.physletb.2017.09.002}{{\em Phys. Lett.}
  {\bfseries B773} (2017) 408--411},
\href{http://arxiv.org/abs/1705.08856}{{\ttfamily arXiv:1705.08856 [nucl-ex]}}.
%%CITATION = ARXIV:1705.08856;%%.

\bibitem{Brun:1997pa}
R.~Brun and F.~Rademakers, ``{ROOT: An object oriented data analysis
  framework},''
\href{http://dx.doi.org/10.1016/S0168-9002(97)00048-X}{{\em Nucl. Instrum.
  Meth.} {\bfseries A389} (1997) 81--86}.
%%CITATION = NUIMA,A389,81;%%.

\bibitem{Tsukada:2017llu}
K.~Tsukada {\em et~al.}, ``{First elastic electron scattering from $^{132}$Xe
  at the SCRIT facility},''
  \href{http://dx.doi.org/10.1103/PhysRevLett.118.262501}{{\em Phys. Rev.
  Lett.} {\bfseries 118} no.~26, (2017) 262501},
\href{http://arxiv.org/abs/1703.04278}{{\ttfamily arXiv:1703.04278 [nucl-ex]}}.
%%CITATION = ARXIV:1703.04278;%%.

\end{thebibliography}\endgroup
\ifprint
\bibliographystyle{ieeetr}
\else
\bibliographystyle{utphys}
\fi
%%%%%%%%%%%%%%%%%%%%%%%%%%%%%%%%%%%%%%%%%%%%%%%%%%%%% 
\appendix 
%%%%%%%%%%%%%%%%%%%%%%%%%%%%%%%%%%%%%%%%%%%%%%%%%%%%% 
\section{Comparison with optical Glauber}
\label{app:opt}
\begin{figure}[t!]
  \centering
  \includegraphics[width=0.45\textwidth]{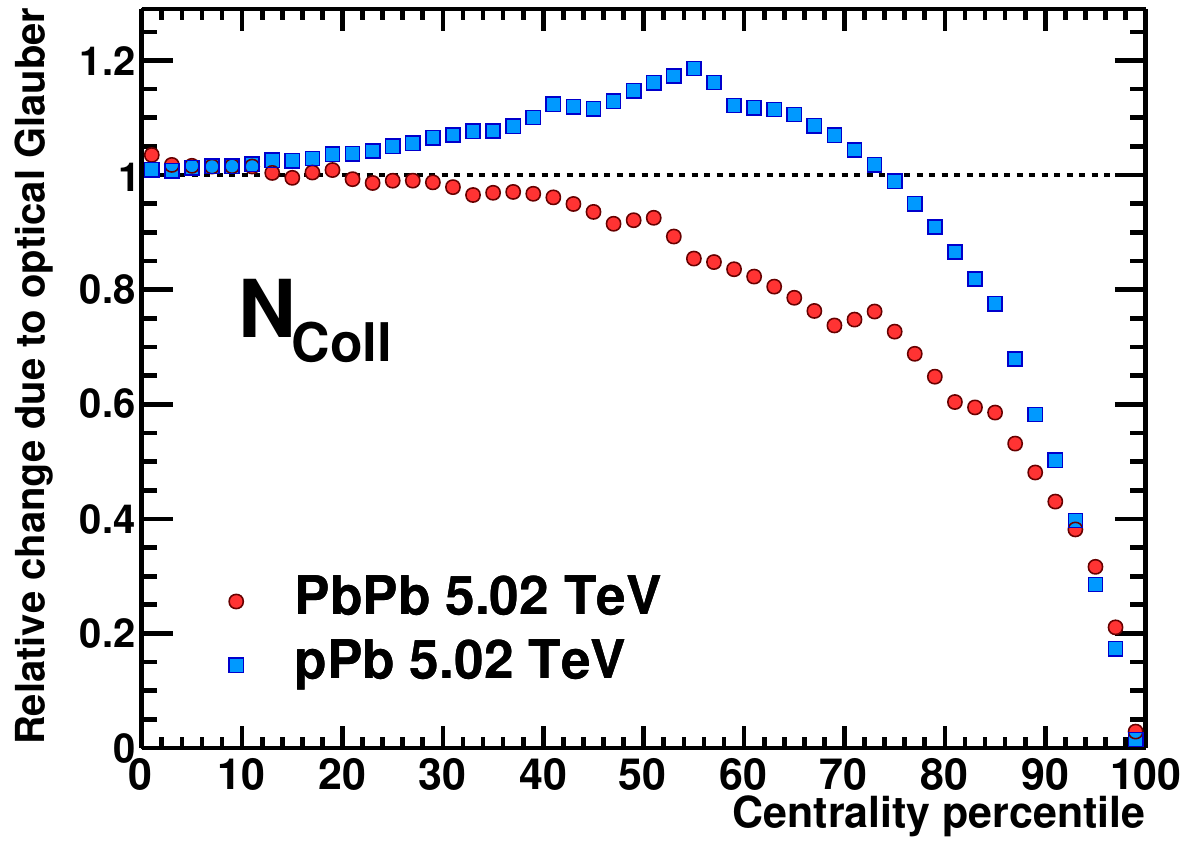} %from macros/mc/DrawNCollPanels_cl.C
  \caption{Relative change in \Ncoll\ in \PbPb\ and \pPb\ collisions at $\snn=5.02$~TeV due to the use of optical Glauber model. The baseline uses the traditional \MCG\ approach with $\sigmaNN=70$~mb.}
  \label{fig:NCollChangeOpt}
\end{figure}

%%%%%%%%%%%%%%%%%%%%%%%%%%%%%%%%%%%%%%%%%%%%%%%%%%%%% 
As described in \Sec{sec:glauber}, the underlying Glauber formalism is the same for optical and MC calculations.
Nevertheless, as discussed in \Ref{Miller:2007ri}, there are differences in their results, in particular in peripheral collisions.
This is demonstrated in \Fig{fig:NCollChangeOpt}, where $\Ncoll$ in peripheral collisions deviates strongly between an optical and the Monte Carlo Glauber calculation.
The optical Glauber calculation~\cite{overlap} was performed with the same parameters for the 2pF distribution of Pb. 
Similarly, the proton was described in the same way, namely with an exponential distribution $\exp\left(-r/R\right)$ with $R=0.234$~fm based on the measured form factor of the proton~\cite{Hofstadter:1956qs}.
For peripheral \PbPb\ collisions beyond 60\% centrality the two calculations differ by more than 20\%, and in the case of \pPb\ collisions the ratio is even non-monotonous.
Optical calculations, which assume a smooth matter distribution, and by construction cannot capture event-by-event fluctuations, should not be trusted in this regime.

%%%%%%%%%%%%%%%%%%%%%%%%%%%%%%%%%%%%%%%%%%%%%%%%%%%%% 
\section{Subnucleonic degrees of freedom}
\label{app:nnmod}
%%%%%%%%%%%%%%%%%%%%%%%%%%%%%%%%%%%%%%%%%%%%%%%%%%%%% 
Potential improvements can be added to the \MCG\ model to take into account subnucleonic dynamics in the nuclear collision, by adding parton degrees of freedom~\cite{Loizides:2016djv}, or fluctuations in the nucleon shape (also known as Glauber-Gribov fluctuations)~\cite{Alvioli:2013vk}. The TGlauberMC includes the possibility that \pp\ collisions themselves have an impact parameter dependence, as \eg\ regularly taken into account in the PYTHIA event generator~\cite{Corke:2011yy}.
A convenient way to include the $\bNN$ dependence in \MCG\ models is to replace the hard-sphere collision condition, $P(\bNN)=\Theta(D-\bNN)$ from \Eq{eq:mc_collisions}, with 
\begin{equation}
 P(\bNN) = \Gamma\left(1/\omega,\frac{\bNN^2}{D^2\omega}\right)/\Gamma\left(1/\omega\right)
  \label{eq:mcwithb}
\end{equation}
where $\bNN$ is the difference between two nucleon centers in the transverse plane, $\Gamma$ is the Gamma function,
and $\omega$ a parameter which covers from the hard-sphere ($\omega = 0$) to the Gaussian ($\omega = 1$) limits.

\begin{figure}[t!]
  \centering
  \includegraphics[width=0.45\textwidth]{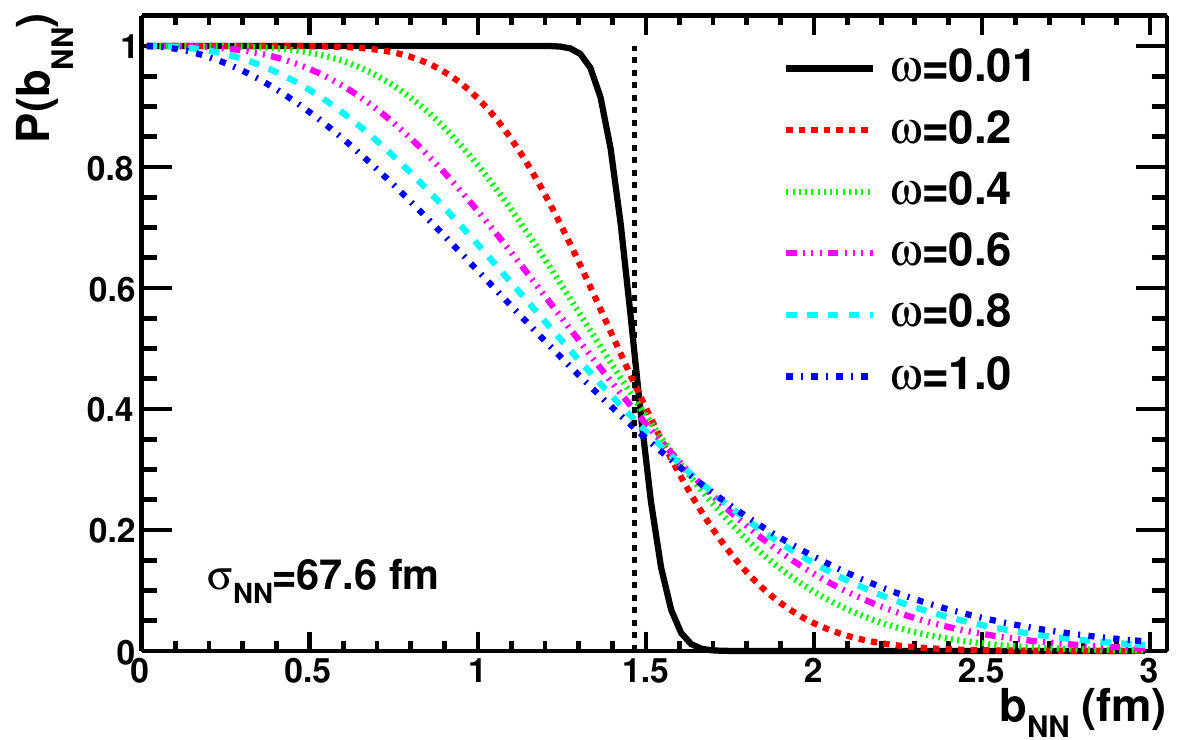} %from macros/mc/plotomega.C
  \caption{Nucleon--nucleon collision impact parameter dependence $P(\bNN)$ from \Eq{eq:mcwithb} for various values of $\omega$,  at 5.02 TeV. The vertical dashed line ($\omega = 0$) corresponds to the hard-sphere limit.}
  \label{fig:Omega}
\end{figure}

\begin{figure}[t!]
  \centering
  \includegraphics[width=0.45\textwidth]{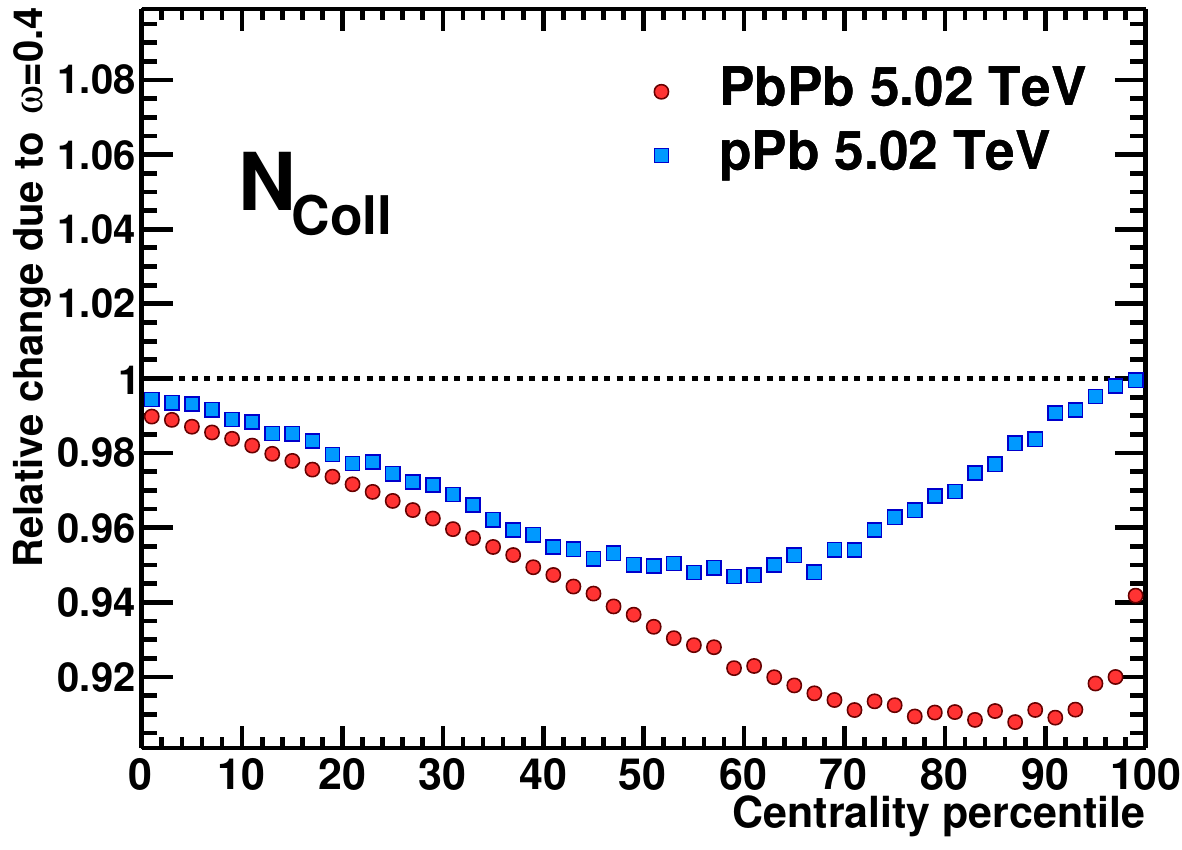} %from macros/mc/DrawNCollPanels_cl.C
  \caption{Relative change in \Ncoll\ in \PbPb\ and \pPb\ collisions at $\snn=5.02$~TeV due to the use of $\omega=0.4$ in \Eq{eq:mcwithb} instead of the hard-sphere condition.
           \co{The baseline uses the traditional approach with $\sigmaNN=70$~mb.}}
  \label{fig:NCollChangeXsecOmega}
\end{figure}

As can be seen in \Fig{fig:Omega}, the resulting probability distribution approaches the hard-sphere step function for $\omega\rightarrow0$ and a Gaussian for $\omega\rightarrow1$.
The proposed value, $\omega=0.4$, for the collisions at the LHC energies reproduces the measured values of both the total and elastic \pp\ cross sections~\cite{Rybczynski:2013mla,Rybczynski:2013yba}.
Using $\omega=0.4$ leads to an effective reduction of the number of collisions relative to the hard-sphere condition, by about $5$\% and $10$\% in peripheral \pPb\ and \PbPb\ collisions, respectively, as shown in \Fig{fig:NCollChangeXsecOmega}.
Since the \MCG\ calculation uses $P(\bNN)$ to determine whether there is a \NN\ collision, using $\omega>0$ in such calculations  will impact the set of generated nucleus--nucleus collisions, and hence all Glauber quantities will change with respect to the typically applied hard-sphere~($\omega=0$) condition, not only $\Ncoll$.  
%Usually the hard-sphere condition is used in \MCG\ calculations. 
The resulting change in $\Ncoll$ is qualitatively similar to earlier studies~\cite{Jia:2009mq} on the influence of the nucleon--nucleon collision geometry on the determination of the $\RAA$.
However, a realistic modeling of the number of hard collisions per \NN\ collision, and in general of the correlation between soft and hard particle production, is needed to be able to compare experimental data with calculations~\cite{Morsch:2017brb}.


%%%%%%%%%%%%%%%%%%%%%%%%%%%%%%%%%%%%%%%%%%%%%%%%%%%%% 
\section{User's guide}
\label{app:guide}
%%%%%%%%%%%%%%%%%%%%%%%%%%%%%%%%%%%%%%%%%%%%%%%%%%%%% 
The source code, which relies on the ROOT~\cite{Brun:1997pa} framework (version 4.00/08 or higher), can be obtained at the TGlauberMC page on HepForge~\cite{glaucode}.
All functionality is implemented in the macro {\tt runglauber\_vX.Y.C}, where version ``X.Y==3.0'' described here.
For generating events with ${}^{3}$H or ${}^{3}$He, the additional text files called ``h3\_plaintext.dat'' or ``he3\_plaintext.dat'' are needed.
Compiling the code is done as in earlier versions, namely by executing
\begin{verbatim}
root [0] gSystem->Load("libMathMore")
root [1] .L runglauber_3.0.C+
\end{verbatim}
Three classes, {\tt TGlauNucleon}, {\tt TGlauNucleus} and {\tt TGlauberMC} and three functions~(macros) {\tt runAndSaveNtuple()}, {\tt runAndSaveNucleons()}, and {\tt runAndSmearNtuple()} are defined in the provided macro. 
In the following, we only describe the improved functionality, see \Ref{Loizides:2014vua} for the complete guide.

Executing the program can be steered by the provided {\tt runAndSaveNtuple()} macro that takes the following arguments:
\begin{verbatim}
 Int_t n,          = number of events
 char *sysA        = name of nucleus A
 char *sysB        = name of nuclear B
 Double_t signn    = inelastic pp cross section
 Double_t sigwidth = width of signn
 Double_t mind     = minimum distance
 Double_t omega    = parameter for NN collision
 Double_t noded    = node distance
 char *fname       = file name
\end{verbatim}
The macro will generate $n$ many Monte Carlo events and store event-by-event computed quantities in a ROOT tree, further described below, saved on disk for a given file name.
If no argument for the file name is given, the code will provide it based on values given for the other arguments.
The names for various nuclear profiles are listed in \Tab{tab:awR} and \Tab{tab:awR2}, and for the corresponding reweighted profiles in \Tab{tab:pol2}, and \Tab{tab:pol2d}.
A complete list for other nuclei can be found in \Ref{Loizides:2014vua}.
All implemented cases can also be found in the {\tt TGlauNucleus::Lookup} function in the code.
The value for $\sigmaNN$ is given in mb, and a variety of values for high energy collisions can be found in \Tab{tab:signnvalues}.
In case a positive value for the width of $\sigmaNN$ is given, then Glauber-Gribov fluctuations~(useful for \pA\ collisions studies) will be simulated.
As a default, a minimum separation distance of $\dmin=0.4$~fm is recommended.
If a positive value for the node distance is given, then the nucleons will be placed on a lattice~(HCP, if not otherwise specified).
For values below $1$~fm the results do not depend on the node distance, but is is recommended to use $0.4$~fm for consistency with $\dmin$.
By default no lattice will be used, and the calculation will be identical to version 2 of the code.
If a positive value of $\omega$ is given, as per \Eq{eq:mcwithb}, the determination of the number of \NN\ collisions will use an \NN-dependent impact parameter distribution as shown in \Fig{fig:Omega}. 
Otherwise, by default, the hard-sphere condition is used.

\begin{table}[t!]
\begin{tabular}{l|c|c|c}\hline\hline
\centering
 Nucleus                                   & Name   & $R$~(fm)          &  $a$~(fm) \\
 \hline
 $^{63}$Cu \co{\cite{DeJager:1987qc}}       & Cu     &  $4.20 \pm 0.02$  &  $0.596 \pm 0.008$ \\
 $^{129}$Xe \co{\cite{Tsukada:2017llu}}     & Xe     &  $5.36 \pm 0.10$  &  $0.590 \pm 0.070$ \\
 $^{197}$Au \co{\cite{DeJager:1987qc}}      & Au     &  $6.38 \pm 0.06$  &  $0.535 \pm 0.027$ \\
\hline\hline
\end{tabular}
\caption{\label{tab:awR2} Nuclear density parameters for charge density distributions of Cu, Xe and Au~(see \Ref{Loizides:2014vua}). 
                          The name of the corresponding profile in the TGlauberMC implementation is also listed. 
                          See \Tab{tab:awR} for Pb.
                          Separate proton/neutron point densities are not known.
                          The values for xenon are obtained from $R=5.4\pm0.1$~fm and $a=0.61^{+0.07}_{-0.09}$~fm for $^{132}$Xe from \Ref{Tsukada:2017llu}, where the radius was scaled down by $0.99 = (129/132)^{1/3}$ and 
                          $a$ was reduced by $0.02$ fm to symmetrize the uncertainty and to approximate the smaller neutron skin of $^{129}$Xe.}
\end{table}

\begin{table}[t]
\begin{tabular}{c|c|c|c|c}\hline\hline
\centering
 Nucleus   & Name     & $p_0$     & $p_1 \cldot 10^4$ & $p_2 \cldot 10^4$\\\hline
 $^{63}$Cu  & Curw     & $1.0090$  & $-7.9040$       & $-3.8990$ \\
 $^{129}$Xe & Xerw     & $1.0091$  & $-7.2230$       & $-2.6630$ \\
 $^{197}$Au & Aurw     & $1.0090$  & $-5.9091$       & $-2.1050$ \\
 $^{207}$Pb & Pbrw     & $1.0086$  & $-4.4808$       & $-2.0587$ \\
 $^{208}$Pb & Pbpnrw   & $1.0087$  & $-4.6148$       & $-2.0357$ \\
\hline\hline
\end{tabular} 
\caption{\label{tab:pol2}Values for the coefficients of the 2nd-order polynomial used to correct the radial nuclear density distribution to cancel the effects of the recentering.
         The name of the corresponding profile in the TGlauberMC implementation is also listed. 
         In case one of the reweighted parameterizations is chosen, the code will by default only generate events with $\dmax<0.1$~fm.}
\end{table}

\begin{table}[t!]
\begin{tabular}{c|c|c|c|c|c}\hline\hline
\centering
 Name     & $\beta_2$   & $\beta_4$   & $p_0$     & $p_1 \cldot 10^3$ & $p_2 \cldot 10^5$\\\hline
 Cu2rw    & $0.162$     & $-0.006$    & $1.0127$  & $-2.9808$       & $-9.9722$ \\
 Xe2arw   & $0.18$      & $0$         & $1.0125$  & $-2.4851$       & $-5.7246$ \\
 Au2rw    & $-0.131$    & $-0.031$    & $1.0126$  & $-2.2552$       & $-3.7151$ \\
\hline\hline
\end{tabular} 
\caption{\label{tab:pol2d}Same as \Tab{tab:pol2} for deformed nuclear profiles corresponding to density parameters given in \Tab{tab:awR2}.}
\end{table}

\begin{figure}[t!]
  \centering
  \includegraphics[width=0.45\textwidth]{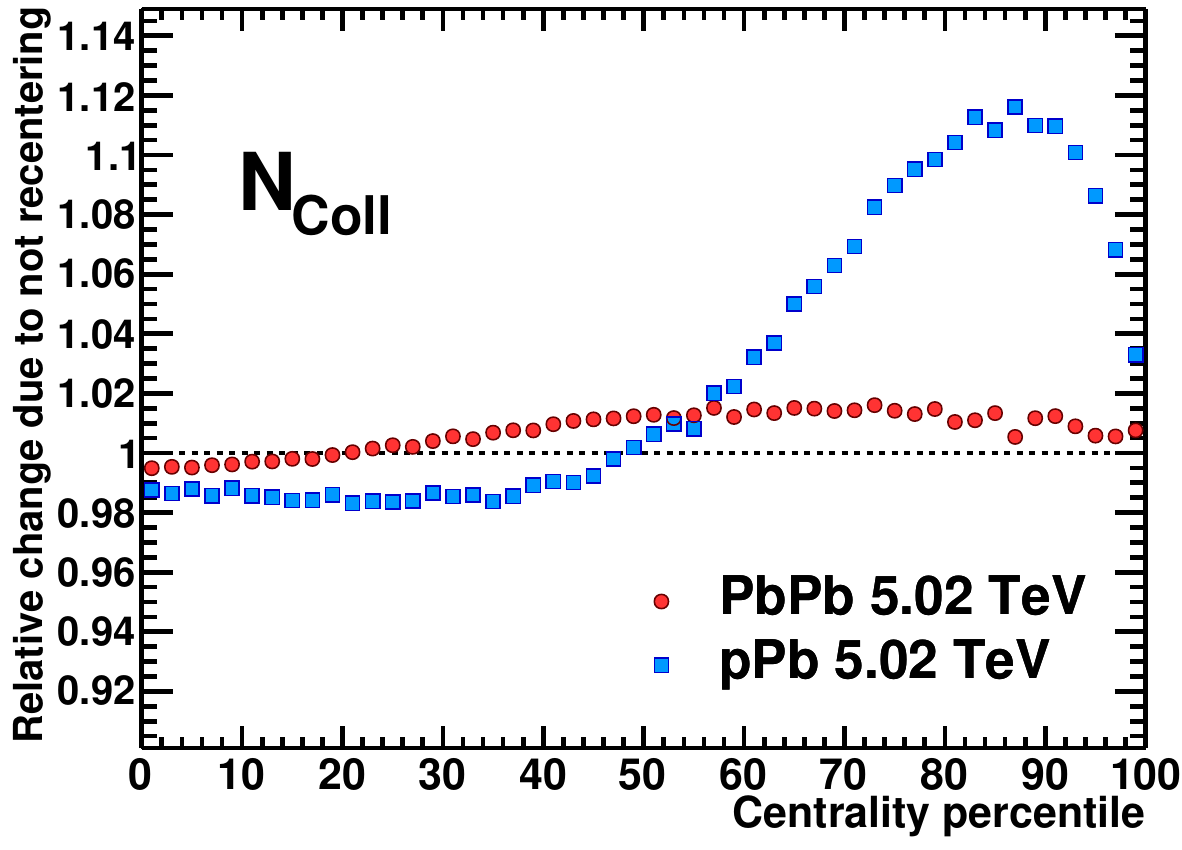} %from macros/tb/DrawNCollPanels_cl.C
  \caption{Relative change in \Ncoll\ in \PbPb\ and \pPb\ collisions at $\snn=5.02$~TeV for $\dmin=0.4$~fm without recentering compared to recentering, using the traditional \MCG\ implementation with $\dmin=0.4$~fm.}
  \label{fig:NCollChangeNorec}
\end{figure}

In addition to quantities described in \Ref{Loizides:2014vua}, the following quantities are stored in the ROOT tree:
\begin{itemize}
\item{\tt Nhard}:    Number of hard collisions (based on fHardFrac)
\item{\tt Ncollpp}:  Number of \pp\ collisions
\item{\tt Ncollpn}:  Number of \pn\ collisions
\item{\tt Ncollnn}:  Number of \nn\ collisions
\item{\tt AreaW}:    Area defined by width of participants
\item{\tt AreaO}:    Area by "or" of participants in grid
\item{\tt AreaA}:    Area by "and" of participants in grid
\item{\tt X0}:       Production point in x
\item{\tt Y0}:       Production point in y
\item{\tt Phi0}:     Direction in $\phi$
\item{\tt Length}:   Length in $\phi_0$      
\end{itemize}

The following set of functions controls additional behavior of the {\tt TGlauberMC} class:
{\tt SetHardFrac(Double\_t)} sets the fraction of cross section used for the calculation of hard collisions (by default $0.65$).
{\tt SetCalcArea(Bool\_t)} and {\tt SetCalcLength(Bool\_t)} enable the calculation of the overlap area using a fine grid and the length (starting from a randomly chosen binary collision with ($x_0,y_0$) in a random direction $\phi_0$ of the transverse plane).
They are by default not computed since the calculation is rather slow.
{\tt SetRecenter(Int\_t)} specifies if and how to recenter nucleons in a nucleus, where 0 means no recentering, 1~(default) means recentering by shifting all nucleons by the average displacement, 2 means recentering by shifting only one nucleon, and 3 recentering by shifting only along the $z$-direction after rotating the nucleus to align the $x$ and $y$ coordinates of its center with 0.
\Figure{fig:NCollChangeNorec} demonstrates the relative change of \Ncoll\ in \PbPb\ and \pPb\ collisions at $\snn=5.02$~TeV for $\dmin=0.4$~fm without recentering compared to recentering using the traditional \MCG\ implementation with $\dmin=0.4$~fm.
{\tt SetShiftMax(Double\_t)} specifies the maximum displacement ($\dmax$) of the nucleon center-of-mass in every direction from zero~(by default any shift is accepted).
{\tt SetLattice(Int\_t)} specifies the lattice type to use~(HCP by default), and {\tt SetSmearing(Double\_t)} specifies the width of a Gaussian by which the nucleon position will be smeared around the lattice node position (by default not done).
{\tt SetBMin(Double\_t)} and {\tt SetBMax(Double\_t)} can be used to restrict the impact parameter (by default between $0$ and $20$~fm).
{\tt SetDetail(Int\_t)} allows one to restrict the number of variables written to the ROOT tree (by default everything is written).
{\tt SetMinDistance(Double\_t)} defines the minimum separation distance (by default $0.4$~fm).
{\tt SetNodeDistance(Double\_t)} sets the node separation in the lattice mode. 
This value should be as large as $\dmin$. By default it is negative, \ie\ the lattice mode is not enabled.
Using {\tt SetNNProf(TF1 *)} one can set another profile than that defined by \Eq{eq:mcwithb}. 
See the code how it is done for {\tt getNNProf}.

\onecolumngrid

%%%%%%%%%%%%%%%%%%%%%%%%%%%%%%%%%%%%%%%%%%%%%%%%%%%%% 
\section{Tables for all computed MCG quantities\\ in $5$\%-wide centrality classes}
\label{app:tab}
%%%%%%%%%%%%%%%%%%%%%%%%%%%%%%%%%%%%%%%%%%%%%%%%%%%%% 
In the following, we present the results for $\Ncoll$, $\Npart$, $\TAA$, $\varepsilon_2$, $\varepsilon_3$, $\AT$, and $L$ in $5$\%-wide centrality classes for all systems, summarized in \Tab{tab:ssum}.
The centrality classes are defined by slicing the impact parameter ($b$) distribution.
For all systems at least 5M events were computed.
For each quantity, the average and the standard deviation~(labeled as rms) of the quantity in the centrality class are reported.
The settings for the improved \MCG\ model are given in \Tab{tab:settings}.

\begin{table*}[b]{\large \PbPb\ at $\snn=2.76$~TeV}

\caption{\label{tab:ccucu02d}Various quantities for \CuCu\ collisions at $\snn=0.2$~TeV obtained with the improved \MCG\ model for centrality classes defined by slicing the impact parameter ($b$) distribution.
         The mean and standard deviation of each quantity~(denoted as \rms) are given. Deformed profile for Cu was used.}
\end{table*}

%%%%%%%%%%%%%%%%%%%%%%%%%%%%%%%%%%%%%%%%%%%%%%%%%%%%%%%%%%%%%%%%%%%%%%%%%%%%%%%
\end{document}